\documentclass[useAMS,usenatbib]{mn2e}
\usepackage{mathtext}
\usepackage{amsmath}
\usepackage{amssymb}
\usepackage{graphicx}
\usepackage{euscript}
\title{Monotonic and cyclic components of radio pulsars spin-down}

\author[A. Biryukov, G. Beskin, S. Karpov]{A. Biryukov$^{1}$\thanks{E-mail:
eman@sai.msu.ru}, G. Beskin$^{2}$\thanks{E-mail:
    beskin@sao.ru}, S.
  Karpov$^{2}$\\
  $^{1}$Sternberg Astronomical Institute of MSU, 13 Universitetsky pr., Moscow,
  119992, Russia\\
  $^{2}$Special Astrophysical Observatory of RAS, Nizhnij Arkhyz,
  Karachaevo-Cherkesia, 369167, Russia\\
}

\begin{document}
\date{Accepted . Received ; in original form }

\pagerange{\pageref{firstpage}--\pageref{lastpage}} \pubyear{201?}

\maketitle

\label{firstpage}

\begin{abstract}

  In this article we revise the problem of anomalous values of pulsars' braking
  indices $n_{obs}$ and frequency second derivatives $\ddot\nu$ arising in
  observations. The intrinsic evolutionary braking is buried deep under
  superimposed irregular processes, that prevent direct estimations of its
  parameters for the majority of pulsars. We re-analyze the distribution of
  ``ordinary'' radio pulsars on a $\ddot\nu-\dot\nu$, $\ddot\nu-\nu$,
  $\dot\nu-\nu$ and $n_{obs}-\tau_{ch}$ diagrams assuming their spin-down to be
  the superposition of a ``true'' monotonous term and a symmetric oscillatory
  term. We demonstrate that their effects may be clearly separated using
  simple {\it ad hoc} arguments.  Using maximum likelihood estimator we derive
  the
  parameters of both components. We find characteristic timescales of such
  oscillations to be of the order of $10^3-10^4$ years, while its amplitudes are
  large enough to modulate the observed spin-down rate up to 0.5-5 times and
  completely dominate the second frequency derivatives. On the other hand,
  pulsars' secular evolution is consistent with classical magnetodipolar model
  with braking index $n\approx3$

  So, observed pulsars' characteristic ages (and similar estimators that depend
  on the observed $\dot\nu$) are also affected by long term cyclic process
   and differ up to 0.5-5 times from their monotonous values. This fact
  naturally resolves the discrepancy of characteristic and independently
  estimated physical ages of several objects, as well as explains very large,
  up to $10^8$ years, characteristic ages of some pulsars.


  We discuss the possible physical connection of long term oscillation with a
  complex neutron star rotation relative its magnetic axis due to influence of
  the near-field part of magnetodipolar torque.
\end{abstract}


\section{Introduction}
\label{sec_intro}

Radio pulsars are highly periodic variable astrophysical objects, powered by
neutron stars' rotation with periods evolving with time.
Observational determination of their timing parameters -- instantaneous period
and its derivatives -- is an extremely complex task
\citep{hobbs_2006_tempo2,cordes_2010}, but after the correction of pulse times
of arrivals (TOAs) for the effects of radio wave propagation in the
interstellar medium, motion and position of the Earth, gravitational delays in
Solar system potential etc., the resulting phase of the light curve may be well
described by an (infinite) Taylor series
\begin{equation}
  \label{eq_phi_series}
  \phi(t) = \phi_0 + \nu(t - t_0) + \frac12\dot\nu (t - t_0)^2 + \frac16\ddot\nu
(t - t_0)^3 + ...
\end{equation}
dominated by the lower order terms. This may also be expressed in terms of
observed
rotational frequency as
\begin{equation}
  \label{eq_nu_series}
  \nu(t) = \nu_0 + \dot\nu(t - t_0) + \frac12\ddot\nu (t - t_0)^2 +
\frac16\dddot\nu (t - t_0)^3 + ...
\end{equation}
where values of $\nu$ and its derivatives are attributed to the epoch $t_0$.

At the same time, the majority of published theoretical models of pulsar
braking predict the spin-down described by a differential equation of the form
\citep{man77, bes93}
\begin{equation}
  \label{eq_dipole}
  \dot\nu=-K\nu^n,
\end{equation}
where $n$ is a (constant) braking index and $K$ depends on the individual
physical properties of the pulsar. For the vacuum magnetodipolar model the
canonical value is $n = 3$, pulsar wind decreases it to $n = 1$ and multipole
magnetic field increases it to $n \ge 5$ \citep{man77}.

For such a spin-down with {\it constant} $K$ the braking index is equal
to a
simple combination of frequency and two its derivatives
\footnote{Note that the \textit{observed} ones, estimated by fitting
  the TOAs with a model of (\ref{eq_phi_series}) broken down to the
third order
  term, are always biased even if the spin-down is exactly according to
  equation (\ref{eq_dipole}) due to influence of non-zero higher-order terms.
For actual
  pulsars, however, this bias is negligible and will be ignored below.}
\begin{equation}
  \label{eq_bi}
  n_{obs} = \frac{\nu \ddot \nu}{\dot \nu^2} = n
\end{equation}
That is why the determination of spin frequency second derivative
via pulsar timing is important for understanding the pulsar' spin-down.

Several decades of very detailed timing of hundreds of pulsars
\citep{d'a93,bay99,chu03,hob04,hob10,liv05}, including the best studied one --
Crab
pulsar \citep{sco03} -- demonstrated, however, that the measured rotational
phase (\ref{eq_phi_series}) evolution is not generally consistent with the one
expected for a braking law (\ref{eq_dipole}) with physically reasonable
parameters. Observed rotational phase $\phi$ (as well as $\nu$, $\dot\nu$)
includes components with a very complex, irregular behaviour.  The first kind of such
irregularities is ``glitches'' -- sporadic fast changes of pulsars' periods
and spin-down rate with up to few months relaxation timescale
(e.g. \cite{man77}), while the other one -- quasi-random phase variations with
typically red Fourier spectrum and characteristic timescales of a few months to
few years -- ``timing noise'' (e.g \cite{cor80,lyn99,hob10}).

Both of them affect the measurements of $\nu$, $\dot\nu$ and $\ddot\nu$, and
often make them dependent on the duration and epoch $t_0$ of the observations'
time span, when it is shorter or comparable to timescales of these processes.
But, if observations are performed over sufficiently long intervals of about a 
decade or longer, the observed coefficients of series (\ref{eq_phi_series}) become
more
stable -- the influence of these irregularities is mostly suppressed (see
Sec.~\ref{subsec_f2data} for additional discussion). To date, there are more
than 200 pulsars observed for longer than 15 years \citep{bay99,
chu03,hob04,
hob10}, and still the majority of $\ddot\nu$ values, measured over such long
intervals, lead to extremely high, up to $10^6$ (and even more), braking
indices. Moreover, their values are negative for nearly a half of all objects. In
general, measured braking indices drastically differ from physically reasonable
values, and this raises serious problem on the pulsars' long term spin evolution.

While glitches are widely believed to be caused by a rapid transfer of momentum
from the neutron stars interior (e.g. \citet{esp11} and references therein), the
nature of the timing noise is still unclear and no self-consistent and widely
accepted model of pulsars' phase residuals irregularities is constructed. There
have been numerous attempts to attribute it to various stochastic phenomena in
the neutron star magnetosphere \citep{cheng_1987,con07,lyn10}, in the interior
\citep{cor85}, to spin-down torque variations
\citep{ura06}, and to the existence of a number of different spin-down regimes
for a single pulsar \citep{lyn10}. Purely phenomenological attempts to describe
observed noise as random walks of different orders (in phase,
frequency or its derivative) \citep{cor81} have also not succeeded as with the
increase of observations time spans the noise appeared to be more complex than
these simple models predict. Some characteristics of this essentially irregular
process, however, have been parametrized through several noise strength
parameters and their correlations. So, \cite{arz94} parametrized pulsars'
timing noise through $\Delta_8 \propto \log(|\ddot\nu|/\nu)$ quantity and found
a good $\Delta_8 - \dot P$ correlation, where $P$ is a pulsar period. 

In turn, \cite{cor80}, \cite{cor85} and then \cite{chu03} have investigated
activity parameter $A$, which is a direct measure of timing residuals variance. A
good correlations of $A$ versus $\dot\nu$ and $\ddot\nu$ was also found. 

However, the quantities used in these papers to parametrize timing noise were
in fact just the measures of observed frequency second derivative. Even the activity
parameter $A$ was calculated using timing residuals after subtraction of a second
order polynomial in series (\ref{eq_phi_series}). Therefore, the correlations
obtained simply reflect the  quite strong $\ddot\nu-\dot\nu$ dependence.


At the same time, only a few long-term mechanisms have been offered to explain
observed anomalous braking indices. \citet{dem79} studied the variations of
pulsar timing parameters due to existence of possible massive partner -- most of
radio pulsars are, however, believed to be isolated objects.  \citet{alp06}
investigated the impact of unresolved or
missed glitches on the observed timing parameters, but it is very difficult to
explain some extremely high braking indices $|n| > 10^2-10^3$ by such
mechanism. \citet{gul77} considered variations of pulsar braking torque with
timescales of $10^2-10^4$ years, probably due to interaction of the pulsar with
interstellar medium in its vicinity. The idea of spin-down torque variations
itself is quite promising and reasonable, but the nature of variations on
such timescale was not pointed out in their note and remained unclear.

In turn, in \citet{bes06} and \citet{bir07} we
considered the existence
of a long-term cyclic variational process affecting pulsars' spin-down on a
timescale of thousands of years. One may call such a process a third type of
timing irregularities. Later, \cite{bar10} suggested that the physical nature of
such a process may be related to the ``anomalous'' magnetodipolar braking
torque, which may produce a forced precession of pulsar's rotational axis
around its magnetic moment with the period of $10^3-10^4$ years, and
demonstrated the possibility of significant variations of observed rotational
parameters on such timescale. Even earlier, \cite{mel00} also investigated the
triaxial neutron star rotation taking into account this torque and
demonstrated that the rotation of such star may be very complex and even induce
variations of magnetic inclination angle, which will also affect the neutron
star spin-down.

As a further development of this concept of a long-term variational process, in
the current work we analyse the ensemble of 297 pulsars with
published data on second frequency derivatives and demonstrate that these
values may be used to estimate the parameters of both secular (evolutionary,
monotonous) and additional, cyclic, components of a pulsar spin-down under
simple and reasonable assumptions. We formulate such a two-component model of
pulsar braking and derive its parameters using a maximum likelihood
estimator. The results are quite reasonable, as the parameters of monotonous
component are in a good agreement with a standard magnetodipolar spin-down
model.

The article is organized as follows. In Section~\ref{sec_data} we justify that
measured $\ddot\nu$ values can indeed be used to analyse pulsars' spin-down. In
Section~\ref{sec_statistics}, statistical analysis of a 297 objects is
performed, phenomenology of a two-component pulsars' spin-down is presented and
parameters of its irregular term are roughly estimated using model-independent
arguments. In Section~\ref{sec_model} this two-component model is used for the
determination of the secular spin-down parameters also. In
Section~\ref{sec_discuss} we discuss the results and its astrophysical
implications. The conclusions are given in Section~\ref{sec_conclude}.

\section{Pulsars' $\ddot\nu$ measurements}
\label{sec_data}

\subsection{What $\ddot\nu$'s can tell us on the pulsars' physics?}
\label{subsec_f2data}

Anomalous values of $\ddot\nu$ (and braking indices) for most of ordinary
pulsars raise at least two principal questions, critical for any serious
analysis of these quantities. Namely:
\begin{itemize}
\item Are values measured over long timespans (decades) really related to
pulsars' or interstellar medium\footnote{The effects of propagation of pulsar
radiation through its ``near'' and ``far'' medium can introduce an additional
delays to the times of pulses arrivals and therefore affect the timing
solution \citep{cor02,cordes_2010}.} physics or they are just artifacts of
somewhat incorrect observations
and/or data reduction?
\end{itemize}
and
\begin{itemize}
\item If they are not artifacts, then is the anomaly induced by the same
phenomena responsible for the short time scale (months) quasi-periodic timing
irregularities with red-like spectra, observed in timing residuals? 
\end{itemize}

The answer to the first question above is clear.  Pulsar timing is a complex,
but
well-defined and thoroughly described procedure, and the sources of
observational uncertainties arising in it are well known and are taken into
account during the reduction, as well as various effects that affect the timing
systematically -- e.g. motion of the Earth, relativistic effects, etc (see, for
example, \cite{hobbs_2006_tempo2}). There is no reason to believe the
significant cubic trends, clearly seen in phase solutions for hundreds of
pulsars \citep{hob04}, are artifacts. 

There is also a method for indirect measurements of a pulsar' second derivative
by comparing $\dot\nu$ values acquired over sufficiently distant, separated by 
$6-20$ years, time spans \citep{jo99}. Results of such estimation are
anomalous too, and are generally consistent with directly measured values.

Obviously, all accurately measured $\ddot\nu$ values, even being anomalous, are
physically meaningful -- they are indeed due to peculiarities of pulsars' spin
down, while their anomalousness is just an indication of insufficiency of our
models of pulsars' braking.


On the other hand, the answer to the second question is not so obvious, as
stochastic timing irregularities, seen in a large fraction of pulsars, still are 
not well understood and are complex phenomena. 


Phenomenologically, timing residuals' irregularities and unexpected anomalous parameters of
timing solution have to be separated in the analysis.  Indeed, the former ones
are seen directly within observational datasets and their stochastic nature is
clear while the origin of the latter ones can only be speculated about. Below
we argue in favour of indeed different origins of these phenomena.

The phase residuals after all appropriate corrections and quadratic trend
subtraction (i.e. after extraction of $\nu$ and $\dot\nu$ only) form a very
plural zoo over the pulsar population. There are at least several different classes of
typical pulsars' timing series (see Figure 3 in \cite{hob10}) -- the ones
with dominant influence of a cubic term (e.g. PSRs B0959-54, B0943+10, B1657-13
etc.), the ones with more complex but generally smooth behaviour (PSRs
B1620-26, B1706-16, B1727-47 etc.), and the ones that are purely white noise
supposedly due to measurement errors.
The latter
class typically contains
pulsars with inaccurate or unmeasured second frequency derivatives, and
will not be analysed here. The second one is often used to depict the timing
deviations from an expected spin-down law as a process with red noise-like
spectrum; there are, however, some evidences that this behaviour in fact is a
combination of a cubic order term and a quasi-periodic component
\citep{hob10}. Moreover, young pulsars (e.g. Crab, B1509-58, B2011+38) show
strong short-term quasi-periodic timing noise, along with cubic trends
corresponding to a spin-down with physically meaningful braking indices $\sim3$
\citep{lyn93, sco03, liv05, hob10}.

Also, the behaviour of $\ddot\nu$'s with the increase of observational span is
not consistent with their origin due to stochastic or short timescale noise
processes. For example, for PSR B1706-16, variations of $\ddot \nu$ with an
amplitude of $10^{-24}$ s$^{-3}$ have been detected on a timescale of several
years (see Figure~7 in \citet{hob04}) -- its values, however, are always around
the one revealed by the fit over the entire 25 year time span
($\ddot\nu=3.8\times10^{-25}$ s$^{-3}$), which still leads to a braking index
$\approx 2.7\times10^3$.

All this strongly suggest that timing noise is a process distinct from the
one producing large (and anomalous) cubic trends in ~\eqref{eq_phi_series};
this noise acts mostly as a randomizing factor decreasing the accuracy of
$\ddot\nu$ measurements, but not defining their properties in general.

As a result, we believe the second derivative values  to be very
promising tool for studying long timescale features of pulsars'
spin-down. Although the $\ddot\nu$ values of individual pulsars bring little
information on the spin-down of neutron stars, the properties of its
distribution over an {\it ensemble} of all pulsars seems to be appropriate for
spin-down study under a simple and reasonable assumptions on the properties of
process producing these anomalous cubic trends.





\subsection{The subset}
\label{subsec_subset}

The set of pulsars under investigation is similar to
the one used in our previous works \citep{bes06, bir07}. From the 393 objects
of the ATNF\footnote{{\tt http://www.atnf.csiro.au/research/pulsar/psrcat/},
revision from 6th Oct 2007}
catalogue \citep{man05} with known $\ddot \nu$ we have compiled a list of
``ordinary'' radio pulsars that
\begin{itemize}
\item have $P>20$ ms and $|\dot P| > 10^{-17}$ s/s (i.e. had not been recycled);
\item have relative accuracy of second derivative measurements better than
  75\%;
\item are not components of binary systems and not in a list of known anomalous
  x-ray pulsars.
\end{itemize}
19 supplementary pulsars from other sources \citep{d'a93,chu03} have been added.
The final set consists of 297 objects including 247 from \citep{hob04}. 18 of
them are associated with young supernova remnants\footnote{Data on PSR-SNR
associations was also taken from ATNF.}

\section{Statistical analysis of pulsars' timing parameters}
\label{sec_statistics}

\begin{figure*}
  {\centering \resizebox*{1.7\columnwidth}{!}{\includegraphics[angle=270]
      {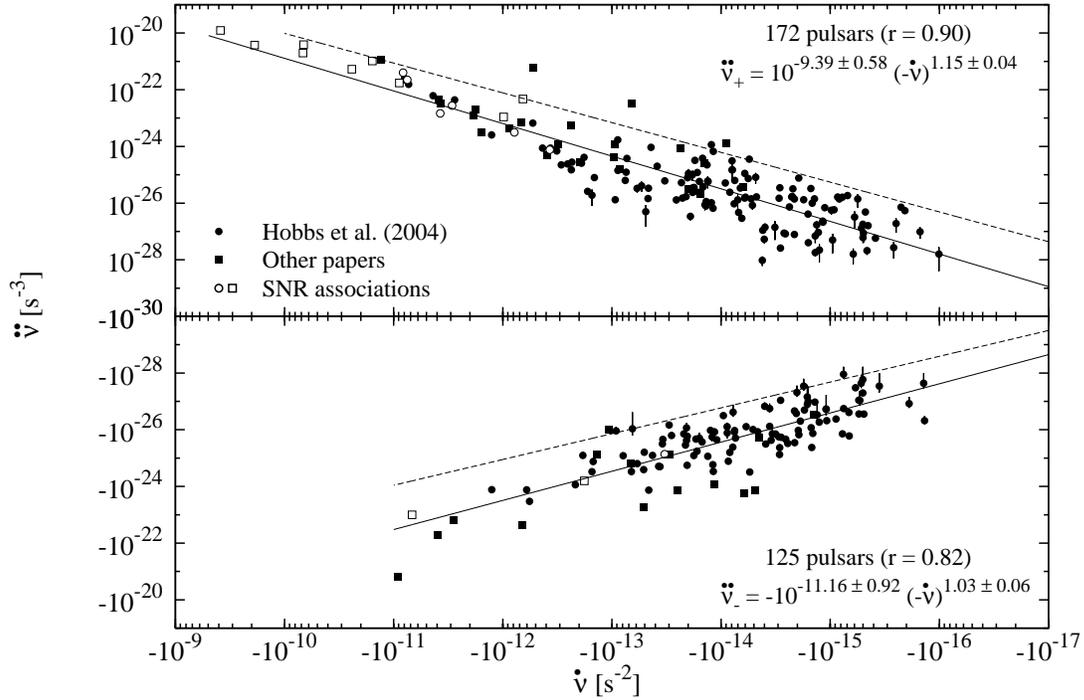}} \par}
  \caption{The $\ddot \nu - \dot \nu$ diagram for 297 pulsars. The figure shows
    the pulsars taken from \citet{hob04} as circles, and the objects measured
    by other groups as squares. Open symbols represent relatively young pulsars
    confidently associated with supernova remnants.  Analytical fits for both
    positive and negative branches are shown as solid lines. They were obtained
    by means of linear least-squares regression in the logarithmic scale.
    Measurement errors are shown as error bars and in most cases are well
    inside the symbols. The diagram is an evolutionary sequence -- pulsars
    systematically move from left to right of it. Two branches are due to
    cyclic variations of pulsars' rotational parameters on a timescale
    significantly shorter than lifetime but longer than time spans of
    observations \citep{bir07} as illustrated in Figure~\ref{fig_sketch}. The
    dashed lines represent the power-law
    approximations of the upper envelopes of positive and negative branches,
    estimated by the method described in \citep{car09}. Their moduli may be
    used as an upper and lower limits on the variational amplitude of $\ddot\nu$,
    respectively (see Section~\ref{sec_limits}).}
  \label{fig_diagram}
\end{figure*}

\begin{figure*}
  \vspace{-4cm}
  {\centering \resizebox*{1.7\columnwidth}{!}{\includegraphics[angle=270]
  {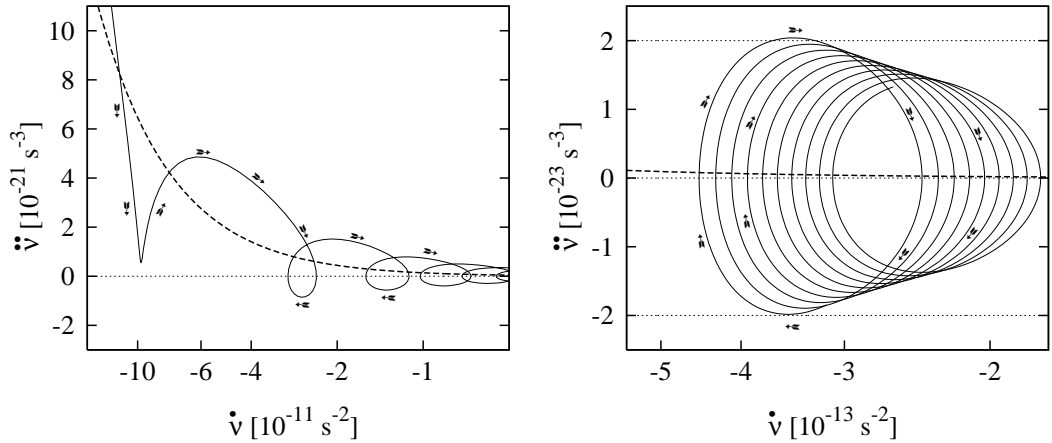}} \par}
\caption{Concept illustration of a pulsar motion on the $\ddot\nu-\dot\nu$
  diagram due to a cyclic evolution of the timing parameters. Monotonous
  component
  of a spin-down (dashed line) is according to a canonical law
  $\dot\nu_{ev} = -K\nu_{ev}^3$ with $K = 10^{-12}$ s$^2$ and
  $\nu_{ev}(0) = 30$ Hz. Superimposed cyclic spin-down component is of a
  $\delta\dot\nu(t) = \dot\nu_{ev}(t)A\cos(\Omega t)$ form, with $A = 0.2$ and
  $2\pi/\Omega = 10^3$ years. Resultant track of a pulsar is shown as the solid
  line. On the {\it left} panel, an initial stages of pulsar evolution are
  presented -- the track is shown for $10^{10}-10^{11}$ s interval of
  ages. Here the amplitude of $\ddot\nu$ variations is not much different from
  evolutionary value $\ddot\nu_{ev}$, and therefore the youngest pulsars are
  unable to change the sign of their observed $\ddot\nu$. On the
  {\it right} panel the evolution of the same pulsar during
  $(1-1.3)\times 10^{12}$ s ages interval is shown. Here, $\ddot\nu$
  variations prevail, and the pulsar repeatedly changes the sign of
  $\ddot\nu$. Note, that due to positive contribution of $\ddot\nu_{ev}$,
  variations are never exactly symmetric in respect to $\ddot\nu = 0$.}
  \label{fig_sketch}
\end{figure*}

\begin{figure*}
  {\centering
    \resizebox*{1.7\columnwidth}{!}{\includegraphics[angle=270]
      {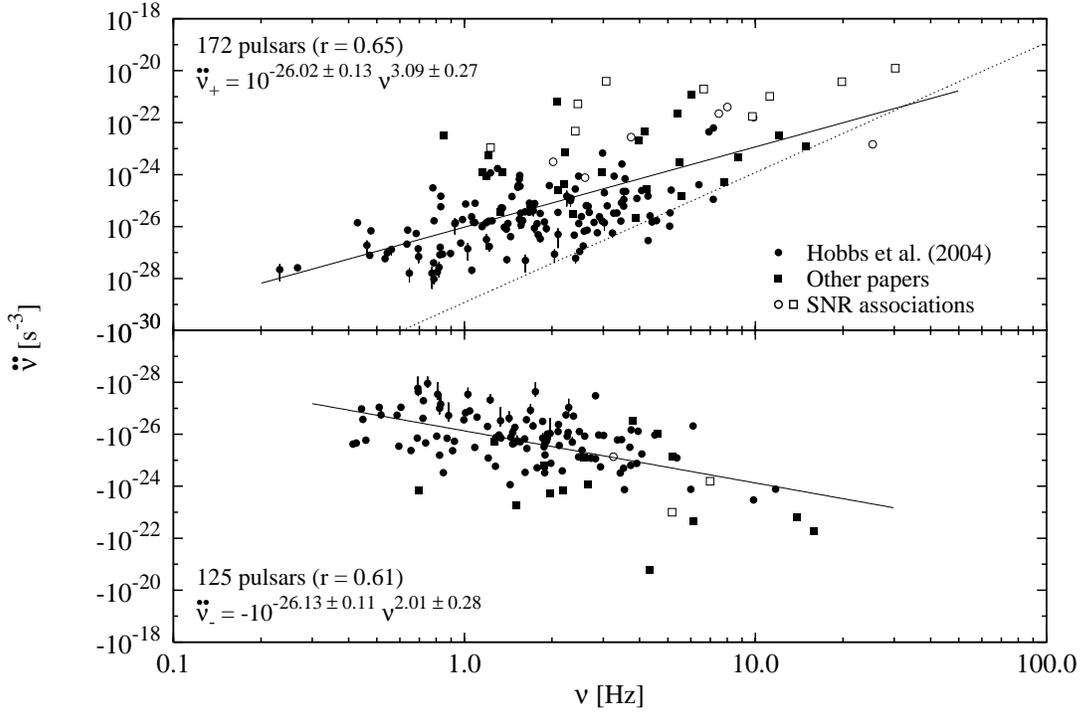}} \par}
  \caption{$\ddot\nu-\nu$ diagram for the subset of pulsars under
    investigation. This diagram is generally similar to the $\ddot\nu-\dot\nu$
    diagram
    shown in Fig.~\ref{fig_diagram} and has the same evolutionary meaning, as
    pulsars born with higher values of $\nu$ and evolve to the lower
    ones. Dotted line shows typical evolutionary trend for a pulsar evolving
    according to $\dot\nu = -K \nu^n$ law with $K = 2\times10^{-15}$ and
    $n = 3$.}
  \label{fig_f2f0}
\end{figure*}

\begin{figure*}
  {\centering \resizebox*{1.7\columnwidth}{!}{\includegraphics[angle=270]
      {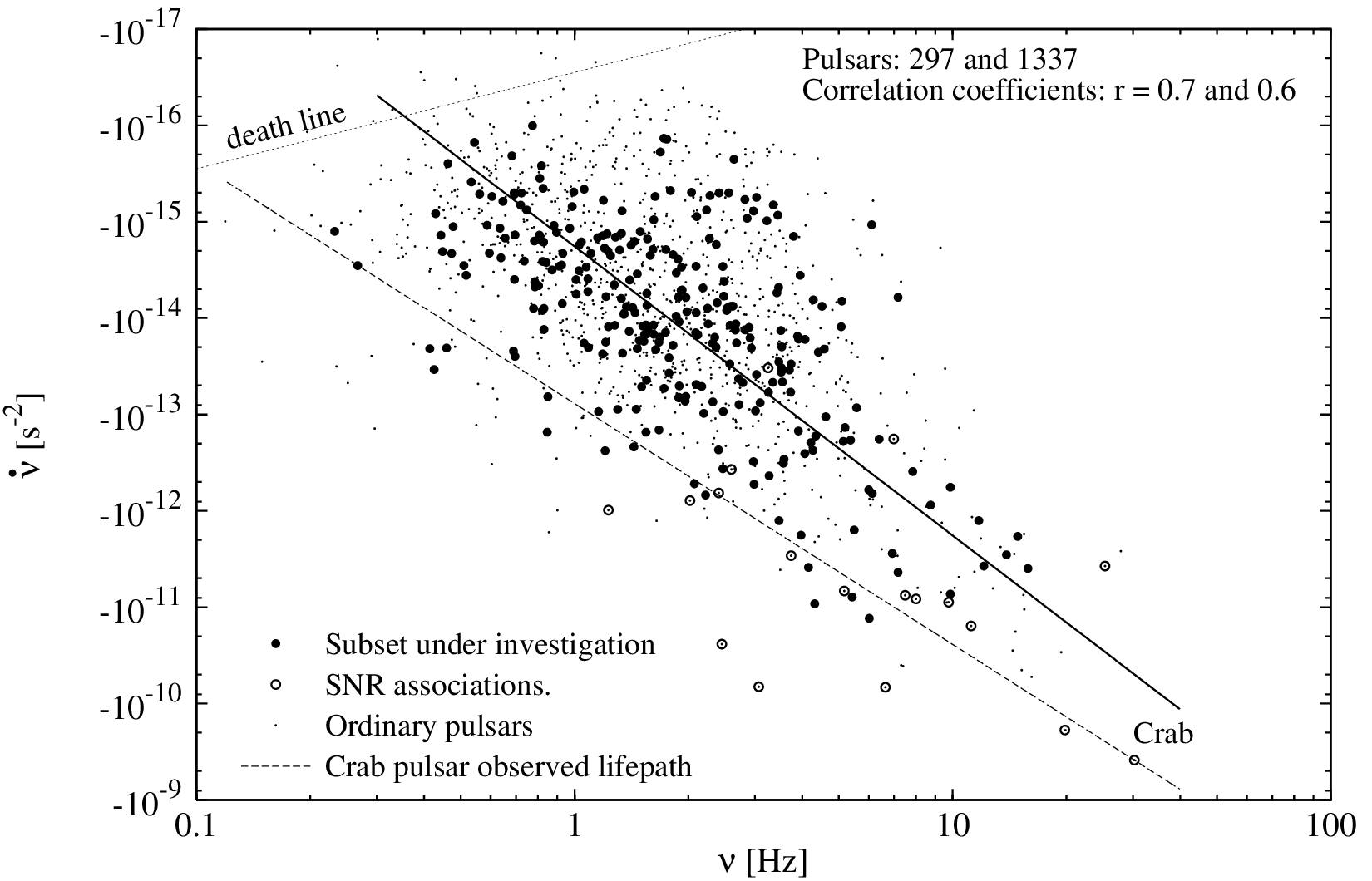}} \par}
  \caption{$\nu-\dot\nu$ diagram for 297 pulsars under investigation (circles)
    and 1337 ``ordinary'' isolated radiopulsars from ATNF database that
satisfy the 
    criteria from Sec.~\ref{subsec_subset} excluding $\ddot\nu$
    measurements.  Extrapolation of present Crab pulsar motion assuming
    simple power-law spin-down with $n=2.5$ is shown as dashed line.  Also,
the
    solid line shows a similar power-law trend for a typical pulsar with
    parameters derived through maximum likelihood estimator
    (Sec.~\ref{sec_model}) -- i.e. with $n=3$ and
    $K = \hat K(n=3,\langle A \rangle = 0.65, \sigma[A] =0.1)$. The pulsars'
    death line $\dot\nu = -2.82\times10^{-17}\nu^{-1}$ is also shown according
    to \citep{bha92}}
  \label{fig_f0f1}
\end{figure*}

\subsection{$\nu-\dot\nu-\ddot\nu$ dependency. Cyclic evolution of pulsars'
spin-down?}
Our work is based on the study of relations between quantities derived
from pulsars' timing: spin frequency, its two derivatives $\dot\nu$ and
$\ddot\nu$, observed braking index $n_{obs} = \ddot\nu\nu/\dot\nu^2$ and
characteristic age $\tau_{ch} = -\nu/2\dot\nu$.

As we previously demonstrated in \citep{bir07}, the logarithms of $|\ddot\nu|$
show significant correlation with the ones of $-\dot\nu$ for both positive (172
pulsars, correlation coefficient $r\approx0.9$) and negative (125 pulsars,
$r\approx0.82$) branches of the $\ddot\nu-\dot\nu$ diagram
(Figure~\ref{fig_diagram}).

The apparent separation of branches on the $\ddot\nu - \dot\nu$ diagram is
primarily due to the logarithmic scale of the plot, and actually objects
continuously cover the full range of $\ddot \nu$ values, excluding only the gap
on small ones due to limited accuracy of measurements (which is not better than
$10^{-29}$ $s^{-3}$).

\begin{figure}
  {\centering
    \resizebox*{1\columnwidth}{!}{\includegraphics[angle=270]
      {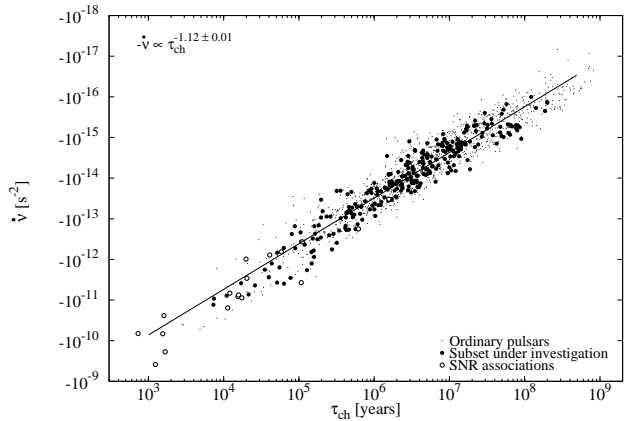}} \par}
  \caption{The correlation between $\dot\nu$ and characteristic age
    $\tau_{ch} = -\nu/2\dot\nu$ for 297 pulsars with measured $\ddot\nu$, as
    well as for 1337 ordinary isolated pulsars taken from ATNF database. The
    correlation coefficient between logarithms of these quantities is
    $r = 0.99$. The spread of points on the scatter plot is due to the
    distribution of frequencies $\nu$ among the pulsars. It, however, is
    unable
    to break tight relation between $\tau_{ch}$ and $\dot\nu$. Therefore, the
    frequency first derivative may also be considered a characteristic of
    pulsar age. Additionally, the obtained dependency
    $-\dot\nu \propto \tau_{ch}^{\alpha}$ with $\alpha \neq -1$ suggests the
    presence of significant $\dot\nu-\nu$ correlation (see
    Fig.~\ref{fig_f0f1})}
  \label{fig_f1_tau}
\end{figure}

Young pulsars confidently associated with supernova remnants are systematically
shifted to the left on the diagram (open symbols in Figure~\ref{fig_diagram})
and are
absent on the right. Hence, pulsars seem to evolve towards
lower values of $|\dot\nu|$. Moreover, if $\tau_{ch}$ is an appropriate
pulsar age estimator, then $|\dot\nu|$ is the same too, as they are highly
correlated ($r = 0.99$ in logarithmic scale, see Fig.~\ref{fig_f1_tau})

On the $\ddot\nu-\nu$ diagram, shown in Figure~\ref{fig_f2f0}, pulsars'
behaviour is similar: they are born with higher values of $\nu$ and,
since $\dot\nu < 0$, evolve toward lower values. The direction of the
evolution
is the same for positive and negative branches of $\ddot\nu-\nu$. So all older
pulsars have systematically lower values of $|\ddot\nu|$ which is consistent
with evolutionary interpretation of $\ddot\nu-\dot\nu$ diagram. 

So, we conclude that each pulsar during its evolution moves along the branches
of the $\ddot\nu-\dot\nu$ and $\ddot\nu-\nu$ diagrams while its $\dot \nu$ value
increases and $\nu$ value decreases. However, on the negative
branch of $\ddot\nu-\dot\nu$, the value of the first derivative, being negative
too,
can only decrease with time, since $\ddot \nu$ is a formal derivative of
$\dot \nu$, and both of them are regular components of the observed rotational
phase (\ref{eq_phi_series}). So, pulsar motion along the branch may only be 
backward, which clearly contradicts its evolutionary interpretation suggested
earlier. The solution we offer is to assume a {\it cyclic} behaviour of
pulsars on this diagram. As pulsars evolve, they repeatedly change sign of
$\ddot \nu$, in a spiral-like motion from branch to branch, and spend roughly
half their lifetime on each one. Such behaviour is sketched in
Figure~\ref{fig_sketch}.


The timescale of such variations has to be much shorter than the pulsar life
time, and at the same time significantly larger than the one of the
observations to systematically affect the timing solution. Of course, such
cyclic behaviour should also affect other spin-down parameters -- $\nu$ and
$\dot\nu$.



Then, on $\dot\nu-\nu$ plot, shown in Figure~\ref{fig_f0f1},
pulsars move from bottom right to the upper left, from high to low $\nu$
and $|\dot\nu|$, forming quasi-evolutionary sequence, with linear regression
coefficient $\sim2$ in logarithmic scale. Unfortunately, this diagram itself
can't be used to study the spin-down evolution of pulsars, as it is distorted
by various selection effects (death line crossing, emission beam width
dependency on pulsar frequency \citep{tau98}, etc) and correlation of pulsars'
initial $\nu$ and $\dot\nu$ at birth.




\subsection{Phenomenology of observed pulsars' spin-down}
\label{subsec_phenom}

Now we introduce an appropriate phenomenological description of the complex
timing evolution of pulsars including the cyclic variations we proposed above.

The exact form of a pulsar's trajectory on the $\ddot\nu-\dot\nu$ diagram is
unknown, but in general its rotational evolution may be described as the
superposition of the regular component of the observed rotational phase
$\phi_{ev}(t)$, and an irregular, {\it cyclic} one $\delta\phi(t)$:
\begin{equation}
  \phi(t) = \phi_{ev}(t) + \delta\phi(t),
  \label{eq_phi_decomp}
\end{equation}
where 
\begin{equation}
  \int_{\Delta T}{\delta\phi(t)dt} \approx 0
\end{equation}
over a long enough time interval $\Delta T$ comparable to a pulsar's
lifetime. After term by term differentiation of equation (\ref{eq_phi_decomp})
one gets:
\begin{equation}
  \nu(t) = \nu_{ev}(t) + \delta\nu(t)
  \label{eq_nu_decomp}
\end{equation}
for the rotational frequency and
\begin{equation}
  \dot\nu(t) = \dot\nu_{ev}(t) + \delta\dot\nu(t) = \dot\nu_{ev}(t)\left[1 +
\varepsilon(t)\right]
  \label{eq_dnu_decomp}
\end{equation}
for the spin-down rate $\dot\nu$. Obviously, due to secular loss of rotational
energy, the value of $\dot\nu_{ev}(t)$ should be always negative for an
isolated pulsar. Quantity
$\varepsilon(t) \equiv \delta\dot\nu(t)/\dot\nu_{ev}(t)$ is a spin-down rate
relative deviation, which is not necessarily small, and is likely greater than
-1:
\begin{equation}
  \varepsilon(t) > -1
\end{equation}
due to the absence of ``ordinary'' pulsars with $\dot\nu > 0$ in the subset
under investigation. It is clear that deviation $\varepsilon(t)$ is also
cyclic; in the simplest case it is a harmonic function of variational phase
$\varphi(t)$:
\begin{equation}
  \label{eq_eps_cos}
  \varepsilon(t) = A \cos \varphi(t),
\end{equation}
where $\varphi(t) \equiv \varphi_0 + \Omega t$ and $A$ is the relative amplitude
of 
the $\dot\nu$ oscillations, related to the absolute amplitude ${\cal
A}_{\dot\nu}$:
\begin{equation}
   {\cal A}_{\dot\nu}(t) \equiv A\dot\nu_{ev}(t)
\end{equation}

By second differentiation of equation (\ref{eq_phi_decomp}) for a
rotational frequency second derivative we get:
\begin{equation}
  \label{eq_ddnu_decomp}
  \ddot\nu(t) = \ddot\nu_{ev}(t) + \delta\ddot\nu(t) = \ddot\nu_{ev}(t)[1 +
  \eta(t)]
\end{equation}
where $|\delta\ddot\nu(t)| \gg \ddot\nu_{ev}(t)$ (i.e. $|\eta(t)| \gg 1$) for
most pulsars, and this is the reason behind anomalously high values
of observed $|\ddot\nu|$ and braking indices. In other words, the absolute
amplitude of the $\ddot\nu$ variations ${\cal A}_{\ddot\nu}$ is much greater
than
the regular term $\ddot\nu_{ev}$. Note also that cyclic terms $\varepsilon(t)$
and $\eta(t)$ are not independent -- they are connected through the relation:
\begin{equation}
  \label{eq_eta_eps}
  \eta(t) = \varepsilon(t) +
\dot\varepsilon(t)\frac{\dot\nu_{ev}(t)}{\ddot\nu_{ev}(t)}
\end{equation}


Since pulsars secularly evolve to the higher values of $\dot\nu$ and
$\dot\nu_{ev}(t) < 0$, the sign of $\ddot\nu_{ev}(t) = d\dot\nu_{ev}(t)/dt$
should always be positive:
\begin{equation}
  \ddot\nu_{ev}(t) > 0
\end{equation}


Such positive contribution should introduce a small asymmetry of
the observed values of $\ddot\nu$ in respect to $\ddot\nu=0$, even if the
variations of $\delta\ddot\nu$ are intrinsically symmetric. Such asymmetry
drives the average motion to the right on the $\ddot\nu-\dot\nu$ diagram and
affects the times that pulsars spend with positive and negative $\ddot\nu$.

\subsection{Asymmetry in observed $\ddot\nu$'s}

Indeed, numbers of pulsars with positive ($N^+ = 172$) and negative
($N^- = 125$) values of $\ddot\nu$ within the subset are significantly
different.  If this difference is just accidental, its probability is too
small, $P = 6.4\times10^{-3}$ only, assuming binomial distribution of 297
objects with $p=1/2$ chance to be on a positive branch. Therefore, null
hypothesis of exactly symmetric branches is rejected on a $0.64$\% significance
level, and the branches are indeed asymmetric in number of
objects\footnote{At the same time, non-parametric Kolmogorov-Smirnov test
  rejects the hypothesis of the common distribution of $|\ddot\nu|$ of
  positive and negative branches on a 2.5\% significance level.}.

\begin{figure}
  {\centering
    \resizebox*{1\columnwidth}{!}{\includegraphics[angle=270]
      {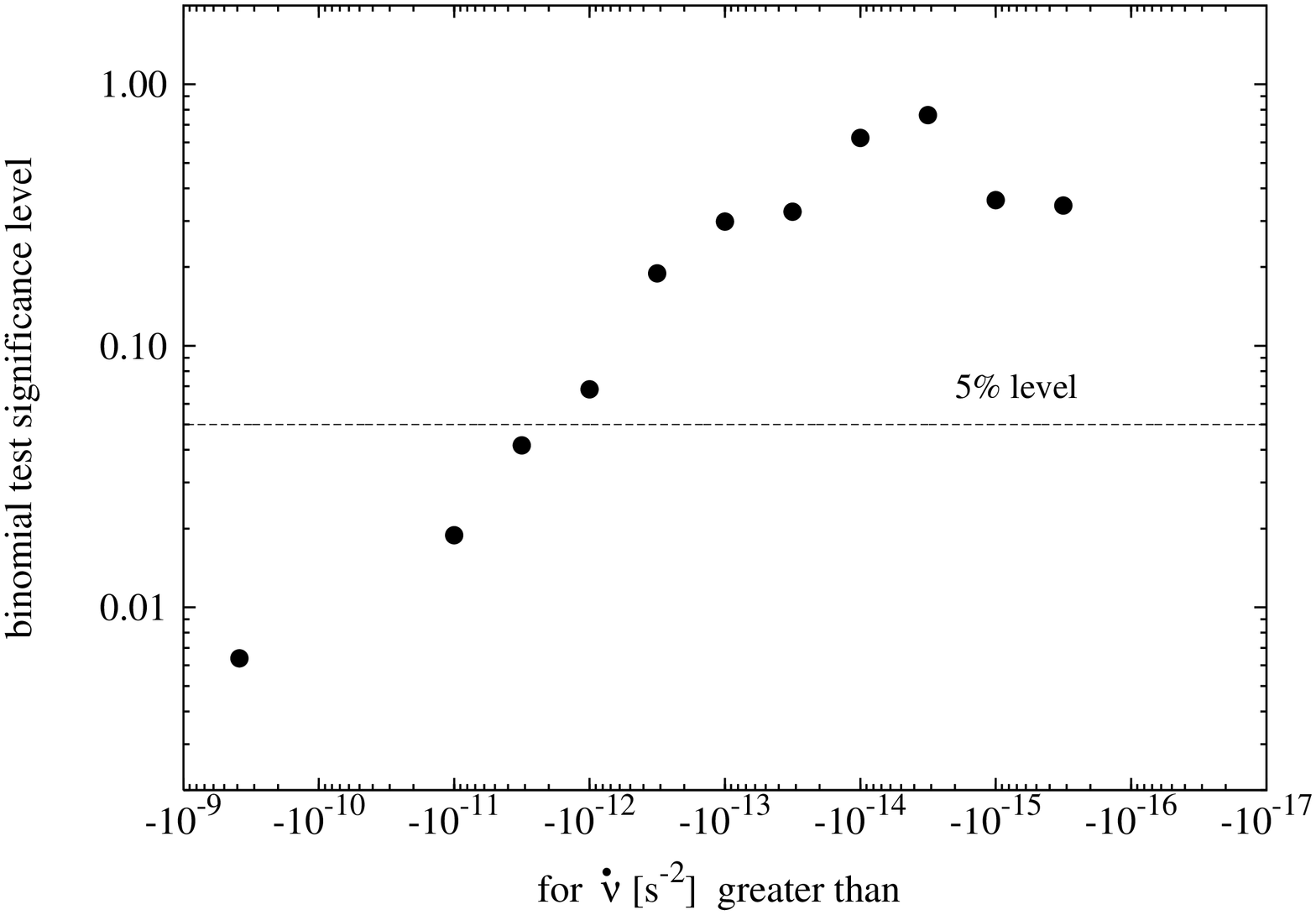}} \par}
  {\centering
    \resizebox*{1\columnwidth}{!}{\includegraphics[angle=270]
      {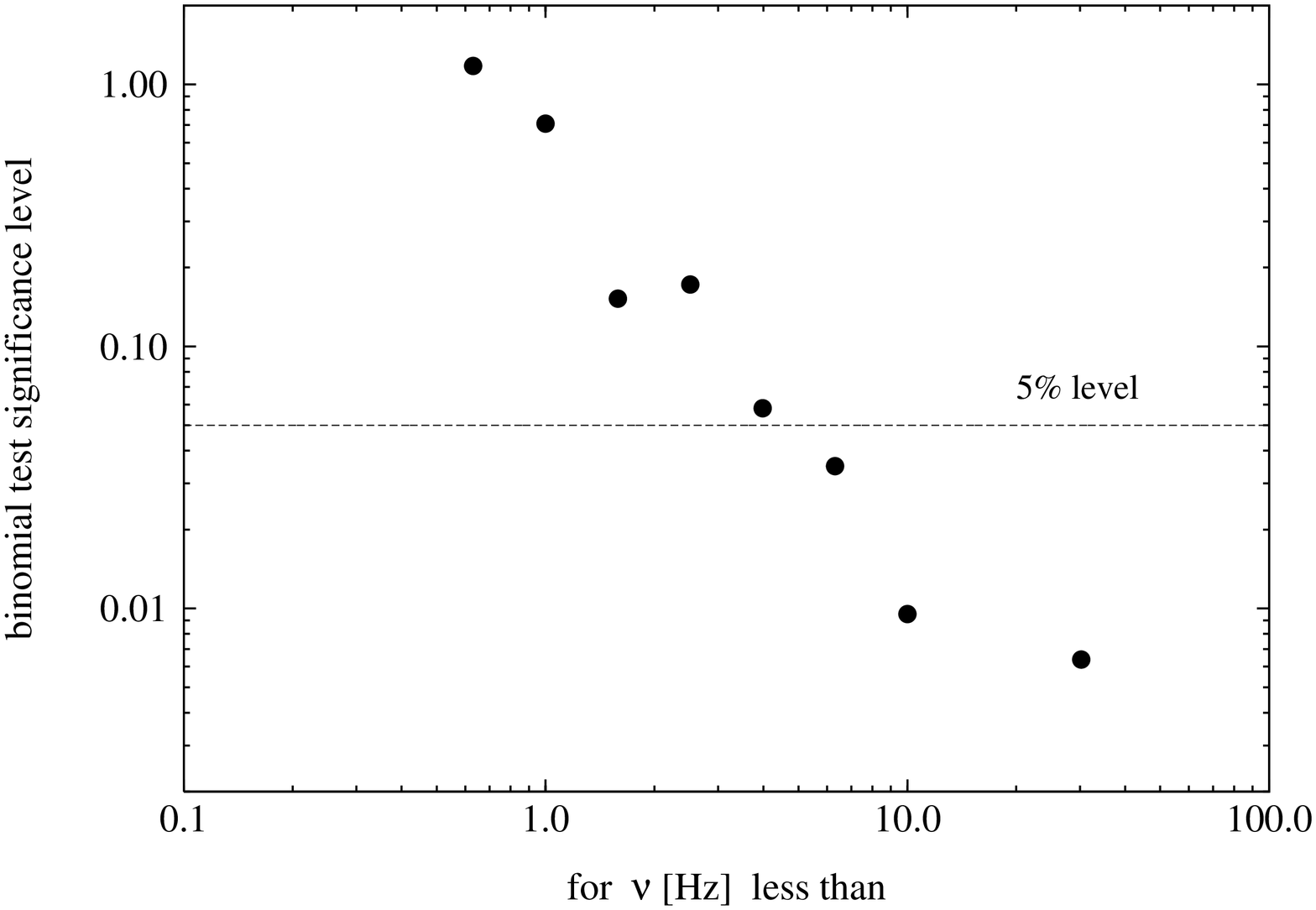}} \par}
  \caption{The significance levels of a non-parametric binomial test for the
    hypothesis of equal numbers of pulsars with $\ddot\nu > 0$ and
    $\ddot\nu < 0$. Each point on the plots represents the $p$-values for the
    subset of pulsars with $\dot\nu$ greater than (upper) and $\nu$ less than
    (bottom) selected threshold. Minimal significance levels correspond to the
    complete set of pulsar. So, the differences of branches in number of
    pulsars are due to contribution of the youngest objects. The absence of
     any
    significant branch differences for older pulsars favors the assumption of
    symmetric variations of $\ddot\nu$ in respect to its evolutionary values.}
  \label{fig_pvalues}
\end{figure}

We investigated in detail the behaviour of such significance levels for a
number of subsets of pulsars with $\dot\nu$ greater than and $\nu$ less then a
given value. Corresponding dependencies are plotted in the
Fig.~\ref{fig_pvalues}. It is clearly seen that branches of the
$\ddot\nu-\dot\nu$ and $\ddot\nu-\nu$ diagrams are significantly different only
when relatively young pulsars are taken into account, while behaviour of the
older pulsars seems to be the same for both signs of $\ddot\nu$. It is
consistent with the existence of a positive evolutionary trend $\ddot\nu_{ev}$,
which is negligible for older pulsars, but is large enough to affect the second
derivatives of the younger ones.

In other words, for the initial stages of pulsars life, values
of $|\delta\ddot\nu|$ are less than $\ddot\nu_{ev}$ (i.e. $|\eta(t)| < 1$), and
the second derivatives are always positive (see the upper left region of
Fig.~\ref{fig_diagram}). Later on, as $|\dot\nu|$ decreases, the
$\ddot\nu_{ev}$ becomes quite small and the relative deviations
$|\eta(t)|$ grow up extremely over unity; as a result, observed values of
$\ddot\nu$ turn out to be significantly different from an intrinsic
$\ddot\nu_{ev}$, and it corresponds to the great increase of their absolute
braking indices (see Fig.~\ref{fig_ntau}). These different stages of pulsars'
evolution are illustrated in the Fig.~\ref{fig_sketch}.

At the same time, the absence of a significant difference between
the $\ddot\nu-\dot\nu$ branches for
relatively old pulsars -- their symmetry -- implies an important fact that
$\delta\ddot\nu$ variations are indeed approximately symmetric in respect to
evolutionary trend: a pulsar deviates to higher and lower values of $\ddot\nu$
with nearly the same amplitude and spends an equal amounts of time with
$\ddot\nu$ greater and less than $\ddot\nu_{ev}$. This fact will be used in
Section~\ref{sec_model} for the estimation of the parameters of pulsars' two-component
spin-down model.

However, some generic characteristics of the cyclic, irregular component of
$\ddot\nu$ behaviour -- its amplitude and timescale -- may be easily derived
from an exploratory analysis of the $\ddot\nu-\dot\nu$ and $n_{obs}-\tau_{ch}$
scatter plots using simple, model-independent arguments. Such analysis is
presented in the next two subsections.

\subsection{Simple limits on the timescales and amplitudes of the cyclic
process}
\label{sec_limits}

As stated above, for the majority of pulsars the amplitude ${\cal A}_{\ddot\nu}$
of the second derivative variations is much greater than the positive
evolutionary trend $\ddot\nu_{ev}$. All the pulsars on the negative branch of
$\ddot\nu-\dot\nu$ diagram are due to subtraction of this
amplitude from the trend. Therefore, the values of $|\ddot\nu|$ of the lower
envelope of this branch (in terms of $|\ddot\nu|$) are sensible estimation of
a {\it lower} limit on ${\cal A}_{\ddot\nu}$. At the same time, the values of
$\ddot\nu$ of pulsars near upper envelope of positive branch may be considered
an {\it upper} limit of ${\cal A}_{\ddot\nu}$.

Indeed, due to existence of the secular trend $\ddot\nu_{ev} > 0$, the amplitude
of the variational term $\delta\ddot\nu$ can not exceed the values of
$\ddot\nu$'s of the upper envelope of the positive branch (for a given $\dot\nu$).
For pulsars there $\ddot\nu = \ddot\nu_{ev} + \delta\ddot\nu =
\ddot\nu_{ev} + {\cal A}_{\ddot\nu} > {\cal A}_{\ddot\nu}$,  while for
ones on the lower envelope of the negative branch $|\ddot\nu| < {\cal
A}_{\ddot\nu} - \ddot\nu_{ev} < {\cal A}_{\ddot\nu}$.


We used the {\tt BoundFit} generalized least-squares code developed by
\cite{car09} 
to acquire an analytical expressions for the mentioned envelopes of the negative
and positive branches on the $\ddot\nu-\dot\nu$ diagram, and found them to be
\begin{equation}
  \label{eq_f2bound_low}
  {\cal A}_{\ddot\nu, low} = 10^{-14.03}(-\dot\nu)^{0.91} \approx
10^{-14.03}(-\dot\nu_{ev})^{0.91}
\end{equation}
and
\begin{equation}
  \label{eq_f2bound}
  {\cal A}_{\ddot\nu, up} = 10^{-9.51}(-\dot\nu)^{1.05} \approx
10^{-9.51}(-\dot\nu_{ev})^{1.05}
\end{equation}
respectively. They are shown as dashed lines in Figure~\ref{fig_diagram}.

Since we consider pulsars on fits (\ref{eq_f2bound_low}) and (\ref{eq_f2bound})
to be observed with amplitude values of frequency second derivative, their
$\ddot\nu$'s are close to turning points, where corresponding $\delta\dot\nu$
change signs during the oscillations (see also the sketch in
Figure~\ref{fig_sketch}), and are obviously small ($\delta\dot\nu \approx
0$). Hence, the assumption of $\dot\nu \approx \dot\nu_{ev}$ in
equations (\ref{eq_f2bound_low}) and (\ref{eq_f2bound}) is justified. 

Due to the absence of pulsars with positive $\dot\nu$, the amplitudes of
variations of frequency first derivatives are likely to be less than their secular
values: ${\cal A}_{\dot\nu} < -\dot\nu_{ev}$ and ${\cal A}_{\dot\nu,up} \sim
-\dot\nu_{ev}$. Also, for any more or less stable
cyclic process, the following relations should be true:
\begin{equation}
 \label{eq_Tdef}
 \Omega \sim \frac{{\cal A}_{\ddot\nu}}{{\cal A}_{\dot\nu}} \sim \frac{{\cal
 A}_{\dot\nu}}{{\cal A}_{\nu}},
\end{equation}
where $\Omega=2\pi/T$ is a characteristic frequency of cyclic variations. (For
the special case (\ref{eq_eps_cos}) these relations are exact). For the 
upper limit $\Omega_{up} \sim {\cal A}_{\ddot\nu,up}/{\cal A}_{\dot\nu,low}$
while for the lower one $\Omega_{low} \sim {\cal A}_{\ddot\nu,low}/{\cal
A}_{\dot\nu,up}$. Hence, one can estimate:
\begin{equation}
 \label{eq_Omega}
 T_{up} = \frac{2\pi}{\Omega_{low}} < - 2\pi\frac{\dot\nu_{ev}}{{\cal
A}_{\ddot\nu, low}},
\end{equation}
which leads to upper limit on the timescale $T_{up}$:
\begin{equation}
 \label{eq_Tup}
 T_{up} \sim (1.2 \times 10^{6}\mbox{ years}) \cdot (-\dot\nu_{ev, 14})^{0.09},
\end{equation}
where $\dot\nu_{ev}$ is normalized to $10^{-14}$ s$^{-2}$. The value of
$T_{up}$ changes from $\approx 8\times 10^5$ years for oldest to
$\approx 2\times 10^6$ years for youngest pulsars. 
Then, the value of lower limit $T_{low}$ should depend on the
upper limit of ${\cal A}_{\ddot\nu}$ and typical minimal relative
$\delta\dot\nu$ amplitude $A_{m}$: ${\cal A}_{\dot\nu,low} = -A_m
\ddot\nu_{ev}$. Hence
\begin{equation}
  \label{eq_Tlowf}
  T_{low} \sim -2\pi \frac{A_{m} \dot\nu_{ev}}{{\cal A}_{\ddot\nu, up}}
\end{equation}
or
\begin{equation}
  \label{eq_Tlow}
  T_{low} \sim (650 \mbox{ years})\cdot A_{m} \cdot
(-\dot\nu_{ev,14})^{-0.05}
\end{equation}
However, $T$ should be more than few times of typical observational span length
for
the pulsars under investigation, which is $15-20$ years. Hence,
\begin{equation}
 \label{eq_Tdown}
 T_{low} \sim 50-100\mbox{ years}
\end{equation}
and $A_{m}$ is unlikely less than $\sim 0.1$. Moreover, we will show below that
$A_m$ is even greater than $\sim 0.5$.

At the same time, assuming that fit to negative branch of $\ddot\nu-\dot\nu$
diagram
\begin{equation}
  \ddot\nu_{-} = -10^{-11.16}\cdot(-\dot\nu)^{1.03} \approx
-10^{-11.16}\cdot(-\dot\nu_{ev})^{1.03}
\end{equation}
represents in the same way a {\it typical} variational amplitude (shown as
solid line $\ddot\nu_-(\dot\nu)$ in Figure~\ref{fig_diagram}), one may, in a
similar way, estimate a typical timescale able to provide observed $\ddot\nu$
spread. So, $\Omega_{typ} = |\ddot\nu_-|/\dot\nu_{ev}$ or
\begin{equation}
 \label{eq_Ttyp}
 T_{typ} = \frac{2\pi}{\Omega_{typ}} = (7.5 \times 10^{4}\mbox{ years}) \cdot
(-\dot\nu_{ev, 14})^{0.03}
\end{equation}
with spread from $\sim 6\times10^4$ to $\sim 9\times10^4$ years for oldest
and youngest objects, respectively.


Finally, according to equation (\ref{eq_Tdef}) the amplitudes of the pulsars'
frequency
variations should be roughly $\Omega^{-1}$ times the ones of its derivative and
$\Omega^{-2}$ times -- its second one. For typical maximal timescale
(\ref{eq_Ttyp}):
\begin{equation}
  {\cal A}_{\nu} < \dot\nu^2_{ev}/|\ddot\nu_-| = (4 \times
10^{-3}\mbox{ Hz})
  \cdot (-\dot\nu_{ev, 14})^{0.97}
\end{equation}
These values do not exceed a few Hertz even for young pulsars and significantly
less than observed frequencies for most of pulsars. It justifies the
neglection of variational term in the rotational frequency decomposition
(\ref{eq_nu_decomp}):
\begin{equation}
  \label{eq_nu_approx}
  \delta\nu(t) < \nu_{ev}(t)\mbox{ and }\nu(t) \approx \nu_{ev}(t)
\end{equation}


\subsection{Pulsar observed braking indices and characteristic ages.
An analysis of the $n_{obs}-\tau_{ch}$ diagram.}
\label{subsec_ntau}

\begin{figure*}
  \vspace{0.5cm}
  {\centering \resizebox*{1.7\columnwidth}{!}{\includegraphics[angle=0]
      {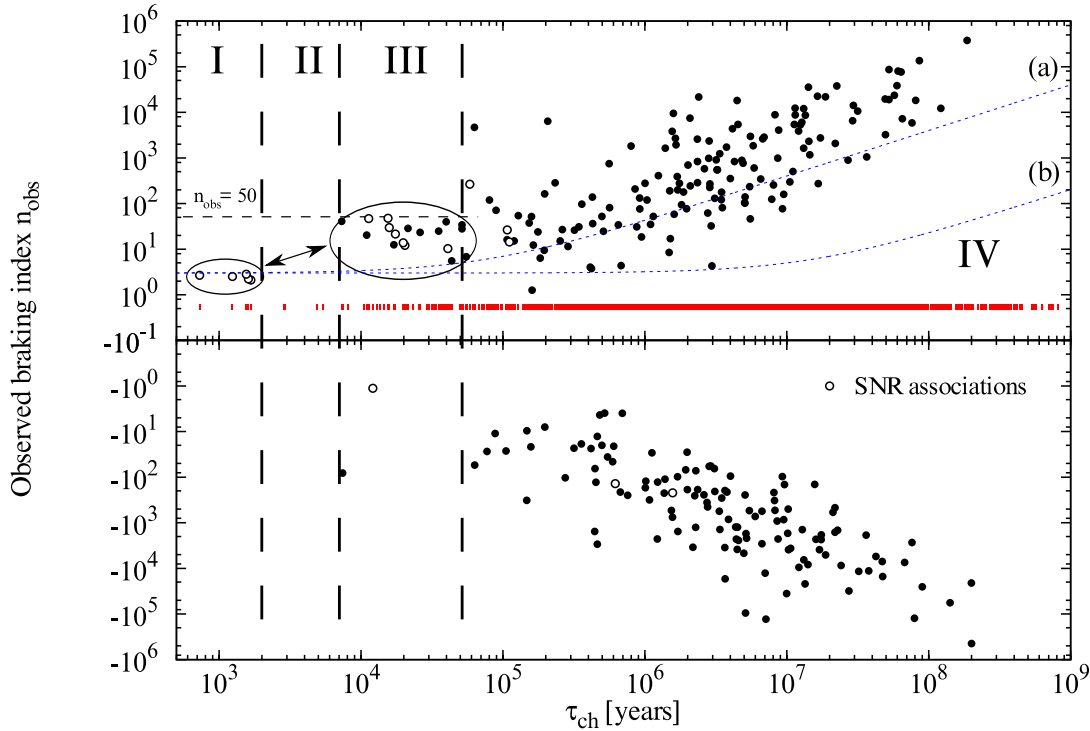}} \par}
  \caption{The $n_{obs} - \tau_{ch}$ diagram. It is similar to the
    $\ddot\nu-\dot\nu$ one shown in Figure~\ref{fig_diagram}. Correlation
    coefficients for positive and negative branches in logarithmic scale are
    0.78 and 0.76 respectively. Red vertical strokes
    represent values of $\tau_{ch}$ of 1337 ordinary pulsars, displayed to
    show overall distribution of $\tau_{ch}$. The younger
    pulsars, particularly associated with SNRs, are located on the left part of
    the diagram (areas I, II and III). Their $\ddot\nu$ amplitudes are still
    quite small and are unable to produce negative observed $n_{obs}$ even in area
    III. However, the braking indices of pulsars there are obviously anomalous
    ($n_{obs}\lesssim 50$), and can not be explained by non-constant
    coefficient in spin-down law (\ref{eq_dipole}) as illustrated by
    blue dashed lines representing pulsar tracks with $-K/\dot K = 5\times10^4$
    (a) and $10^7$ (b) years, respectively. It suggests that
    these values are primarily affected by large $\dot\nu$ variations
    ($A\sim0.5-0.7$). At the same time, the lack of objects in area II
    combined with small $n_{obs}$ of pulsars of area I means that observed
    braking indices and ages of the latter group are likely to be also affected
    by $\dot\nu$ variations. The observed $\tau_{ch}$ of the youngest pulsars
    seem to be less than evolutionary values, while $\tau_{ch}$ of pulsars of
    area III \textit{vice versa} are greater. The amplitudes of $\ddot\nu$
    variations of pulsars of area IV are much greater than evolutionary values
    which produces two nearly symmetric branches of the diagram. }
  \label{fig_ntau}
\end{figure*}

There are number of widely accepted estimators of pulsars' physical properties
based on results of timing solution and power-law (\ref{eq_dipole})
spin-down model with constant $K$ and $n = 3$: characteristic age
$\tau_{ch} = -\nu/2\dot\nu$, rotational energy loss rate $\dot E_{rot}
\propto \nu\dot\nu$ and surface magnetic field
\begin{equation}
  \label{eq_B}
  B = (3.2\times 10^{19}\mbox{ G}) \sqrt{-\dot\nu\nu^{-3}} .
\end{equation}

But, if the amplitude of frequency derivative variations, $|\varepsilon(t)|$,
is large, these quantities may be significantly biased, as they are computed
using observed $\dot\nu$ values, and not actual evolutionary ones $\dot\nu_{ev}$.

%

Using phenomenology introduced earlier and assuming power-law spin-down
defined by equation (\ref{eq_dipole}) with constant $K$ and evolutionary braking
index $n$,
observed quantities $n_{obs}$ and $\tau_{ch}$ may be expressed as:
\begin{equation}
  \label{eq_nobs}
  n_{obs} = n\frac{1 + \eta}{(1 + \varepsilon)^2}
\end{equation}
and
\begin{equation}
  \label{eq_tauobs}
  \tau_{ch} = \left(\frac{n-1}{2}t +
\frac{P^{n-1}_0}{2K}\right)\frac{1}{1 + \varepsilon} =
\frac{\tau_{ch,ev}}{1 + \varepsilon},
\end{equation}
where $P_0$ is pulsar initial period, $t$ -- its real age, and $\varepsilon$
and $\eta$ are relative $\dot\nu$ and $\ddot\nu$ deviations according to
equations (\ref{eq_dnu_decomp}) and (\ref{eq_ddnu_decomp}). If $n = 3$,
then $P^2_0/2K \sim 10^3$ years for the typical pulsar and
$\tau_{ch, ev} \approx t$ for the older objects.

So, if $\varepsilon$ varies in $\pm 0.9$ range, then
half of pulsars will be observed as up to $\tau_{ch}/\tau_{ch,ev}
\approx 1/(1 - 0.9) = 10$ times older than their
true unbiased characteristic ages, while another half as $1/(1 + 0.9) \approx
0.5$ times younger. The effect is therefore seemingly asymmetric, as
characteristic ages will mostly appear overestimated. 

The same bias will also appear in the rotational energy loss rate, while
magnetic fields will be $2-3$ times underestimated for pulsars with $\varepsilon <
0$.

Moreover, since observed braking indices $n_{obs} = \nu\ddot\nu\dot\nu^{-2}
\propto \dot\nu^{-2}$, their absolute values will be additionally increased up
to order of magnitude for a half of pulsars.

Since $n_{obs}$ and $\tau_{ch}$ depend on $\varepsilon(t)$ and $\eta(t)$, the
pulsar behaviour on the $n_{obs}-\tau_{ch}$ plane, shown in
Figure~\ref{fig_ntau}, is similar to that in the $\ddot\nu-\dot\nu$ one.
The positive and negative branches are not identical: the numbers of pulsars
and shapes of branches are significantly different.

The most interesting regions of the diagram are the areas I, II and III with
the relatively young pulsars. There are 19 pulsars with positive and only 2
with negative $n_{obs}$ in the area
III\footnote{Moreover, these two pulsars with negative $\ddot\nu$ (PSR B1338-62
  and PSR B1610-50), while satisfying to the formal selection criteria
  described in Sec.~\ref{subsec_subset}, have relatively bad data sets --
  B1338-62 shows significant glitch activity, and time spans for both of them
  are only about 200 days long.}.
Binomial test with $p = 1/2$ rejects the hypothesis of an accidental origin of
such asymmetry on a 0.01\% significance level. The pulsars with positive
$n_{obs}$ have measured $n_{obs} \sim 10-50$, which are larger than any
reasonable secular value. Moreover, they slightly deviate from the rest of
positive branch of the diagram.

Is it possible to explain these $n_{obs}$ as a result of decay of $K$ 
coefficient in spin-down law (\ref{eq_dipole}), probably due to evolution 
of magnetic inclination angle (e.g. \cite{dav70})? To yield 
pulsars with both $n_{obs} \sim 30-50$ and $\tau_{ch} \sim 10^4$ years,
corresponding timescale $-K/\dot K$ should be as short as few
thousands years only. However, e.g. \cite{fgk06} have shown that there is no
significant decay of coefficient $K$ on a timescale $\sim 10^8$ years for
isolated pulsars.

We plotted pulsar evolutionary tracks corresponding to spin-down law
(\ref{eq_dipole}) with $n = 3$ and $-K/\dot K = 5\times 10^4$ and $10^7$
years on the $n_{obs}-\tau_{ch}$ diagram (blue dashed lines in
Figure~\ref{fig_ntau}). It is clearly seen that first of them is insufficient
to explain $n_{obs}$ of pulsars in area III, while second one is unable to
affect observed distribution of pulsars' $n_{obs}$ at all.

So, we conclude that such asymmetry results from combination of a small
enough amplitudes of $\ddot\nu$ variations (i.e.  $|\eta| \ll 1$, so the
negative values of
$\ddot\nu$ can not be reached yet) and high amplitudes of $\dot\nu$ ones.
Assuming
$n = 3-5$ and $\eta = 0$, for these pulsars with $n_{obs} = 50$ from
relation (\ref{eq_nobs}) one get:
\begin{equation}
  \label{eq_eps_estimate}
  \varepsilon \approx \sqrt{\frac{n}{n_{obs}}} - 1 = -0.76--0.68
\end{equation}
Therefore, the existence of such a remarkable group of pulsars on the
$n_{obs}-\tau_{ch}$ diagram argues in favour of significantly high relative
amplitudes of $\dot\nu$ variations -- $\delta\dot\nu(t)$ in
decomposition (\ref{eq_dnu_decomp}).


Young pulsars in the area III are mostly the ones with negative
$\varepsilon$. Since the variational process is likely symmetric relative
to evolutionary trend $\dot\nu_{ev}$,  young pulsars with positive
values of $\varepsilon$ should exist with $\tau_{ch} < \tau_{ch,ev}$ and
$n_{obs}$ that is likely less than $n$.


The pulsars of area I seem to be exactly such objects. They all have
$n_{obs} < 3$ and their observed characteristic ages are most likely shifted
towards lower values. Also, there is a deficit of pulsars in the area
II\footnote{Note, however, that at least six ``ordinary'' pulsars without
  $\ddot\nu$ measurement are known in this area. Their ages concentrate near
  $\tau_{ch} \sim$ 3 and 5 kyr as seen from the set of red vertical strokes
in Figure~\ref{fig_ntau}}
(with $\tau_{ch} = 2-7$ kyr) which may be a result of pulsars escaping from
there to lower and higher $\tau_{ch}$ due to positive and negative
$\varepsilon$, correspondingly.
So, the variations can be a reason for $n_{obs} < 3$ measured for
the youngest pulsars known. We will return to the analysis of the
youngest pulsars within our concept of cyclic variations in the Discussion.

Finally, all remaining pulsars on the diagram -- objects of the area IV -- form
two nearly symmetric branches with positive and negative $n_{obs}$. The
variational amplitudes of $\ddot\nu$ of these pulsars are much larger than
corresponding evolutionary values: $|\eta| \gg 1$. They are also affected by
large amplitudes of $\dot\nu$ variations that may produce the oldest (with
$\tau_{ch} > 10^7- 10^8$ years) observed pulsars. Indeed, the largest pulsars'
$\tau_{ch}$ seem to be physically unreasonable, as pulsars should cross
their ``death lines'' in a few millions of years only \citep{bha92, fgk06}.

\section{Pulsars' spin-down model}
\label{sec_model}

In the previous section we performed a statistical analysis of pulsars
rotational parameters and concluded that observed pulsar spin-down is
consistent with an idea of the existence of some cyclic variational
process. Model-independent estimations show that this process operates on
timescales as long as few hundreds or thousands of years, while relative
amplitudes of $\dot\nu$ variations are comparable to secular, monotonous
component of pulsar spin-down $\dot\nu_{ev}$.

Also, we found some evidence that $\ddot\nu$ variations are symmetric in
respect to evolutionary term $\ddot\nu_{ev}$. This fact may be useful to
reveal the parameters of the evolutionary component, common for all
pulsars.

In the Section below we construct a two-component model of observed pulsars'
rotational evolution which consists of evolutionary monotonous and cyclic
components according to decomposition (\ref{eq_dnu_decomp}), and derive its
parameters.

\subsection{Monotonous component of observed spin-down}

For the secular spin-down term of our model we will assume canonical power-law
expression:
\begin{equation}
  \label{eq_dnu_ev}
  \dot\nu_{ev} = -K \nu_{ev}^n,
\end{equation}
with $K = const$ unique for each pulsar. On the other hand, we presume the
constant {\it evolutionary} braking index $n$ to be the same for all pulsars.
The observed quantity $n_{obs}$ and model parameter $n$ are not equivalent and
are connected with relation (\ref{eq_nobs}) for each pulsar.

Combining equations (\ref{eq_nu_decomp}), (\ref{eq_dnu_decomp}) and
(\ref{eq_dnu_ev}),
the evolutionary value of second frequency derivative for each pulsar with
observed $\nu$ and $\dot\nu$ may be written either as
\begin{equation}
  \label{eq_f2_f0t}
  \ddot\nu_{ev,1} = n K^2 \nu^{2n - 1}
\end{equation}
or as
\begin{equation}
  \label{eq_f2_f1t}
  \ddot\nu_{ev,2} = n K^\frac{1}{n} \left(-\frac{\dot\nu}{1 + \varepsilon}
\right)^{2 -\frac{1}{n}}
\end{equation}

These expressions correspond to projections of an evolutionary trend onto
either $\ddot\nu-\nu$ or $\ddot\nu-\dot\nu$ planes, respectively. Moreover,
when computed using actually measured timing parameters, these quantities may
be treated as a statistically independent ones, as they arise from combinations
of different observables.

\subsection{Cyclic term of observed spin-down}

As a toy model, not necessarily exact but still representing all the important
properties, we assume below a simple harmonic form for an oscillating term of
sum (\ref{eq_dnu_decomp}):
\begin{equation}
  \label{eq_a}
  \varepsilon(t) = A\cos\varphi(t),
\end{equation}
where $A$ is a constant relative amplitude, and $\varphi(t)$ is a phase of variations.

Actual values of $\varepsilon$ for individual pulsars, as well as values of
coefficient $K$ in the law (\ref{eq_dnu_ev}), are unknown. Therefore, we will
analyse
our subset in terms of its average, ensemble characteristics only. The
parameters of $\varepsilon$'s distribution will be included in the model.

Namely, to take into account physical diversity of individual pulsars, we
assume $A$ to be normally distributed with mean $\langle A \rangle$ and
variance $\sigma[A]$:
\begin{equation}
  \label{eq_distribA}
  A \sim {\cal N}(\langle A\rangle, \sigma^2[A]).
\end{equation}
At the same time, the phase $\varphi$ is obviously distributed uniformly,
\begin{equation}
  \label{eq_distribPHI}
  \varphi \sim Uni(0, 2\pi),
\end{equation}
as it depends (as a modulus of $\varphi = \varphi_0 + f(T, t)$ quantity over
$2\pi$) both on an unknown value of initial phase $\varphi_0$ and randomly
selected moment of observations -- pulsar' age $t$. It is also safe to assume
the phases of individual pulsars to be uncorrelated both over the ensemble and
with pulsar' parameters.

\begin{figure}
  {\resizebox*{1\columnwidth}{!}{\includegraphics[angle=270]
      {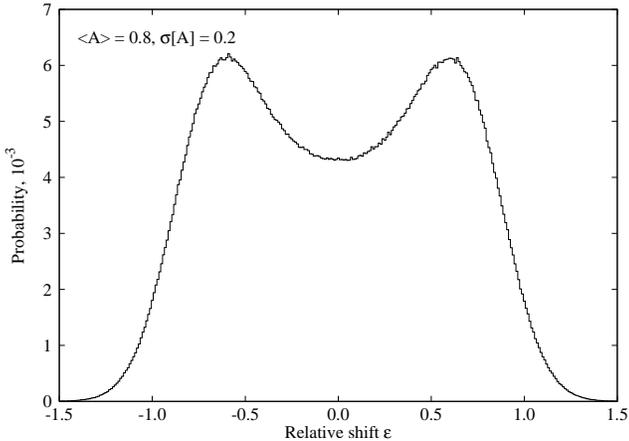}} \par}
  \caption{Distribution of relative deviations $\varepsilon$ simulated
    according to relation (\ref{eq_a}) and distributions (\ref{eq_distribA}) and
(\ref{eq_distribPHI})
    with $\langle A \rangle = 0.8$ and $\sigma[A] = 0.2$. The probability
    $P(\varepsilon < -1)$ determines the probability for pulsar to be observed
    with positive $\dot\nu$.}
  \label{fig_edistrib}
\end{figure}

\begin{figure}
  {\resizebox*{1\columnwidth}{!}{\includegraphics[angle=270]
      {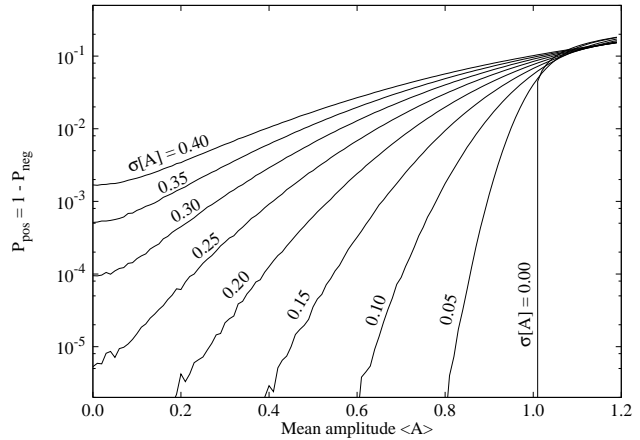}} \par}
  \caption{Probability $P_{pos} \equiv {\cal P}(\varepsilon < -1)$ for a pulsar
    to be observed with
    positive $\dot\nu$ for different combinations of parameters
    $\langle A \rangle$ and $\sigma[A]$ of $\varepsilon$ distribution.}
  \label{fig_ppos}
\end{figure}

Simulated $\varepsilon$ distribution for typical parameters of $\langle A \rangle =
0.8$ and $\sigma[A] = 0.2$ is shown in Figure~\ref{fig_edistrib}.
Values of $\varepsilon$ less than $-1$ correspond to the positive observed
$\dot\nu$ since $\dot\nu = \dot\nu_{ev}(1 + \varepsilon)$. The probability to
find a pulsar with $\dot\nu > 0$:
 \begin{equation}
  \label{eq_ppos}
  P_{pos} \equiv {\cal P}(\varepsilon < -1)
\end{equation}
for such a model depends on the parameters of the $\varepsilon$ distribution
$\langle A\rangle$ and $\sigma[A]$. This dependence is shown in 
Figure~\ref{fig_ppos}.

\subsection{Evolution of an ensemble of pulsars}

Since we assume $n$ to be the same for all pulsars, their trends
$\dot\nu_{ev}(\nu_{ev})$ on the $\dot\nu-\nu$ diagram are characterized by
individual coefficients $K$ and initial frequencies $\nu_0$ only. These trends
are straight lines in logarithmic scale (see Figure~\ref{fig_f0f1}) and do not
intersect each other. Hence, the average behaviour of a pulsars' ensemble may
be characterized by the line separating the set of their evolutionary
trajectories in half. Such trend corresponds to the median value of the
coefficient $K = -\dot\nu_{ev}/\nu^n_{ev}$:
\begin{equation}
  \label{eq_K}
  \hat K = {\cal M}\left[-\frac{\dot\nu}{\nu^n}\frac{1}{1 + \varepsilon}\right],
\end{equation}
where ${\cal M}[\cdot]$ is a median over an ensemble and relations
(\ref{eq_dnu_decomp}) and (\ref{eq_nu_approx}) are used.  The value of
$\hat K = \hat K(n)$ can not be estimated from observables only since values of
$\varepsilon$ for individual pulsars are unknown, but one can build a
distribution of $\hat K$ assuming statistical properties of $\varepsilon$
within the ensemble. This distribution is a function of three parameters:
\begin{equation}
  \hat K = \hat K(\langle A \rangle, \sigma[A], n)
\end{equation}
It will be used below to build a maximum likelihood estimator for the parameters
of the pulsars in the ensemble.

\subsection{Criterion for the estimation of model parameters}

If the $\ddot\nu$ distribution is symmetric with respect to $\ddot\nu_{ev}$,
then the
numbers of objects with observed $\ddot\nu$ less and greater than corresponding
$\ddot\nu_{ev}$ should be nearly equal, both for the whole ensemble and for any
subset defined by quantities uncorrelated with phase (i.e. $\nu$ or
$\dot\nu$). It provides a simple criterion for a model goodness-of-fit for
an appropriately chosen $\ddot\nu_{ev,i}$
(defined by equations (\ref{eq_f2_f0t}) and (\ref{eq_f2_f1t}))
the numbers of pulsars observed with $\ddot\nu > \ddot\nu_{ev,i}$ should
{\it simultaneously}, for both $i=1,2$ be distributed binomially as
\begin{equation}
  \label{eq_nplus_theor}
  N^+_i \sim Bin(N, p=\frac{1}{2}),
\end{equation}
where $N$ is the total number of pulsars in the ensemble. The same should
obviously
be valid for any sub-interval along $\nu$ or $\dot\nu$: if the number of
pulsars in $k$-th sub-interval on $i$-th plane is $N_{ik}$, then the
probability to have $N^+_{ik}$ of $N_{ik}$ pulsars with
$\ddot\nu > \ddot\nu_{ev}$ is
\begin{equation}
  \label{eq_binomial_prob_theor}
   P_{ik} = \frac{N_{ik}!}{(N_{ik} - N^+_{ik})!\cdot N^+_{ik}!} \cdot
   \left(\frac12\right)^{N_{ik}}
\end{equation}

Using this criterion and the dependence of $N^+_{ik}$ on the parameters of
pulsars' spin-down model, an obvious maximum likelihood estimator can be
constructed to derive these parameters.

\subsection{Maximum likelihood estimator for the model parameters}

The estimation of  $N^+_{ik}$ for any interval of $\nu$ or $\dot\nu$
requires exact values of $\ddot\nu_{ev,i}$ and, hence, known specific $K$ and
$\varepsilon$ for each pulsar. These values can not be derived directly from
observational data and therefore we have to hypothesize about them.

We adopted the following routine for its estimation. Since no {\it ad hoc}
information is available on $\varepsilon$, we assumed it to be a random value
distributed according to equations (\ref{eq_a})-(\ref{eq_distribPHI}). The
specific
$K$ has been, in turn, replaced with median $\hat K$ value computed according
to relation (\ref{eq_K}) with these $\varepsilon$ values in mind. This way, the
corresponding
estimates for the monotonous component of the second frequency derivatives for
each
pulsar from equations (\ref{eq_f2_f0t})-(\ref{eq_f2_f1t}),
\begin{equation}
  \label{eq_f2_f0}
  \hat{\ddot\nu}_{ev,1} = n \hat K^2 \nu^{2n - 1}
\end{equation}
and
\begin{equation}
  \label{eq_f2_f1}
  \hat{\ddot\nu}_{ev,2} = n \hat K^\frac{1}{n} \left(-\frac{\dot\nu}{1 +
\varepsilon} \right)^{2 - \frac{1}{n}}
\end{equation}
are in turn random quantities, uncorrelated to each other as they arise from
combinations of different observables. They are, however, median estimates of
its exact values due to functional form of equations, and therefore should
roughly divide an ensemble of pulsars in half for adequately chosen model
parameters -- $n$, $\langle A \rangle$ and $\sigma[A]$.



\begin{figure*}
  {\centering \resizebox*{1.7\columnwidth}{!}{\includegraphics[angle=270]
      {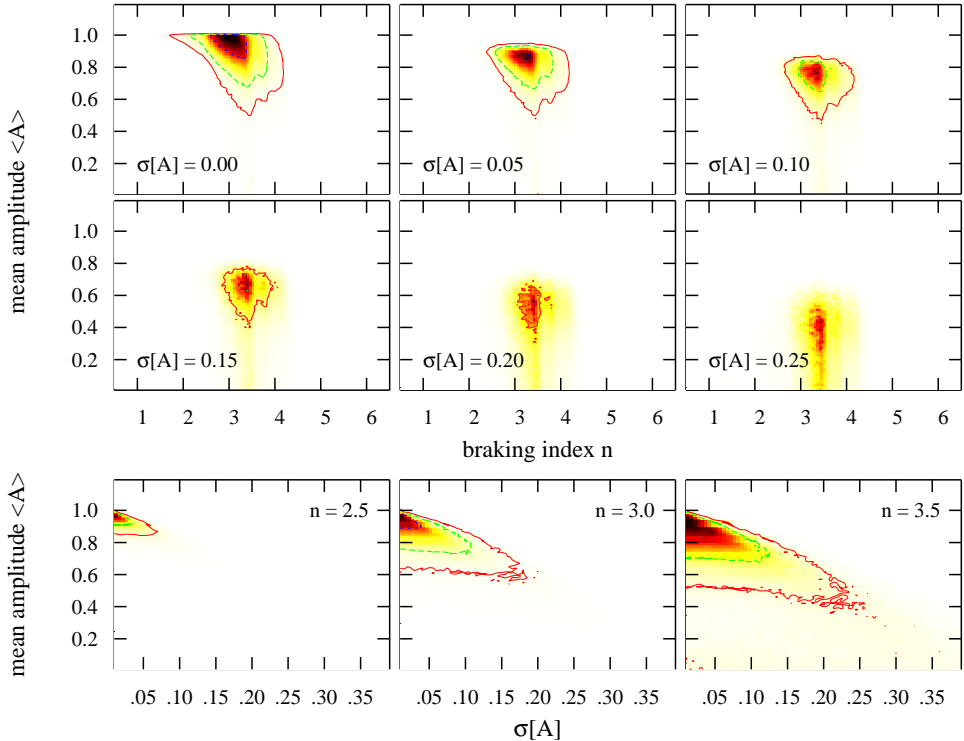}} \par}
  \caption{Plot of the confidence regions derived through the maximum
    likelihood estimator. The number of slices for fixed $\sigma[A]$ and $n$
    are shown. Blue, green and red contours represent 65, 95 and 99\%
    bounds. The values of evolutionary braking index $n$, mean amplitude
    $\langle A \rangle$ and its dispersion $\sigma[A]$, consistent with the
    symmetry of $\ddot\nu$ variations, are at $n \sim 2.5-4$,
    $\langle A \rangle > 0.5$ and $\sigma[A] < 0.25$.}
  \label{fig_slices}
\end{figure*}

So, to construct maximum likelihood estimator we divided all pulsars into $m=6$
intervals of $N_{ik}\approx 50$ objects each ($k = 1,2,...m$) along $\nu$ and
$\dot\nu$ on the $\ddot\nu-\dot\nu$ and $\ddot\nu-\nu$ planes,
respectively. Then we calculated a likelihood function as follows:
\begin{itemize}
\item On the first step specific values of $\langle A \rangle$, $\sigma[A]$ and
  $n$ have been chosen from the parameter space.
\item Then $N = 297$ values of $\varepsilon$ have been simulated according to
distributions (\ref{eq_distribA}) and (\ref{eq_distribPHI}), and
\item $\hat K(n, \langle A \rangle, \sigma[A])$ have been determined.
\item The number of pulsars $N^+_{ik}$ with $\ddot\nu > \hat{\ddot\nu}_{ev,ik}$
in $k^{th}$ interval have been calculated and
\item used to compute binomial probabilities $P_{ik}$
defined (\ref{eq_binomial_prob_theor}).
\item Then, the likelihood functions has been constructed:
\end{itemize}
\begin{equation}
  \label{eq_likelihood0}
  \ell_{i} = \prod_{k = 1}^{m} P_{ik}
\end{equation}
We have also taken into account an absence of pulsars with positive $\dot\nu$
in an ensemble of $N=297$ pulsars. This fact constraints the parameters of 
$\varepsilon$ distribution, as probability $P_{pos}$ for the
pulsar to be observed with positive $\dot\nu$ unlikely exceeds $1/N$ and
corresponding probability
 \begin{equation}
  \label{eq_ps0}
  P_{neg} = 1 - P_{pos} > 1 - \frac{1}{N}
\end{equation}
for the pulsar to be observed with {\it negative} $\dot\nu$ therefore
has to be large enough.

The binomial probability $P_{neg}^N$ to observe all $N$ pulsars with
{\it negative} values of $\dot\nu$ has therefore been added as a multiplicative
term to the likelihood function, and
\begin{itemize}
\item finally we calculated logarithmic likelihood functions as
\end{itemize}
\begin{equation}
  \label{eq_likelihood}
  {\cal L}_i = -N \log P_{neg} -\sum_{k=1}^{m}\log P_{ik}
\end{equation}
The indices $i$ = 1 and 2 here mark (statistically independent) likelihoods for
the planes $\ddot\nu-\nu$ and $\ddot\nu-\dot\nu$, respectively.

We performed simulations of relative deviations
$\varepsilon = \delta\dot\nu/\dot\nu$ for 297 pulsars $10^4$ times in a
row\footnote{All simulations have been performed using the ``Chebyshev''
  supercomputer of Moscow State University.}
for each set of given $\langle A\rangle$, $\sigma[A]$ and $n$.  In this way, the
distributions (but not exact values) of ${\cal L}_i$ for every combination of
parameters have been obtained. We have computed these distributions over a grid
of values $n$, $\langle A\rangle$ and $\sigma[A]$ in the intervals $0.5-6.5$,
$0-1.2$ and $0-0.4$ with steps $0.05$, $0.01$ and $0.05$, respectively.

\subsection{Likelihood for the null hypothesis}

Classical maximum likelihood technique presuppose a search of extremum of
obtained $\cal L$ within parameters space and then constraining the confidence
regions assuming that distribution of $\cal L$ is known if the null hypothesis
is valid. Typically $\cal L$ is a single-valued function of
model parameters, and the $p-$\% confidence region consists of a values of
$\cal L$ for that $P({\cal L}_0 < {\cal L}) \le p/100$, where ${\cal L}_0$ is a
random quantity distributed as a logarithmic likelihood within null
hypothesis. The minimum of $\cal L$ is obviously within this confidence region.

This method may be generalized in a straightforward way for the case when
$\cal L$ is not a deterministic function of parameters but a random quantity
with known distribution.

To do it, we simulated the distribution of ${\cal L}_0$ by computing its values
in a way similar to one used earlier for ${\cal L}_i$, but using $N^+_{ik}$
{\it simulated} according to binomial distribution with $p = 1/2$, and
{\it simulated} uniformly distributed $P_{neg}$ on the $[(N-1)/N;1]$
interval. Therefore, the null hypothesis corresponds to the exact symmetry of
numbers of pulsars with $\ddot\nu$ greater and less than
$\hat{\ddot\nu}_{ev,i}$ in each of $m$ intervals described above.


Then, using known distribution of $\cal L_0$, we independently estimated the
probabilities ${\cal P}_i = {\cal P}({\cal L}_0 < {\cal L}_i)$ and conflated
them into the resulting united value $\cal P$ using the method described by
\cite{hill10} and \cite{hill11}:
\begin{equation}
  \label{eq_conf}
   {\cal P} = \frac{{\cal P}_1 {\cal P}_2}{{\cal P}_1 {\cal P}_2 + (1 - {\cal
P}_1)(1 - {\cal P}_2)}
\end{equation}
This relation represents the cumulative probability that both likelihood
functions have an extremum in the same point of a model parameters space.
Thus, the isocontours of $\cal P$ mark the bounds of a $100\cdot (1-{\cal P})$
percent confidence regions.

\subsection{Results of maximum likelihood estimation}
\label{subsec_results}

The maps of final conflated probabilities $P$ are shown in the
Fig.~\ref{fig_slices} as a collection of slices of the parameter space for
fixed $\sigma[A]$ (upper plots) or $n$ (lower plots). The 99\% confidence
interval covers the range of $n$:
\begin{equation}
  \label{eq_n_est}
  2.5 < n < 4
\end{equation}

The corresponding range of $\langle A \rangle$ depends on the accepted value of
$\sigma[A]$. The decision about the appropriate amplitude and its dispersion
can be made from analysis of $n_{obs}-\tau_{ch}$ diagram
provided above. Indeed, according to estimation (\ref{eq_eps_estimate}), typical
relative deviations $\varepsilon$ are as high as $0.7\div0.8$, which corresponds
to the $\sigma[A] \sim 0.1$ in Figure~\ref{fig_slices}. So, we believe that the
most adequate choice is:
\begin{equation}
  \label{eq_A_est}
  0.5 < \langle A  \rangle < 0.8,
\end{equation}
and
\begin{equation}
  \sigma[A] \sim 0.1
\end{equation}

Obtained values of $n$ are in a good agreement with the $n = 3$ expected from
simple magnetodipolar pulsar braking theory with constant (or slowly evolving)
$K$ \citep{man77}.

The median evolutionary trend, assuming $n = 3$, relative $\dot\nu$ amplitude
$\langle A \rangle = 0.65$ and $\sigma[A] = 0.1$, is shown in
Figure~\ref{fig_f0f1}.

\subsection{Timescales of cyclic process}
\label{sec_timescales}

The model of observed pulsars' spin-down does not include explicit relation
between variational phase $\varphi$, pulsar age and timescale of variations
$T$. To estimate the distribution of $T$, we will formulate it separately.
Namely, let
\begin{equation}
  \varphi = \varphi_0 + \Omega t,
\end{equation}
where $\Omega = 2\pi/T = const$.

Since functions $\varepsilon(t)$ and $\eta(t)$ defined earlier are not fully
independent and related through equation (\ref{eq_eta_eps}),
the timescales (or frequencies) of a variational process for our model may be
estimated directly: assuming secular spin-down law (\ref{eq_dnu_ev}),
$\varepsilon = A\cos\varphi = A\cos(\Omega t + \varphi_0)$,
$\dot\varepsilon = -\Omega A \sin(\Omega t + \varphi_0)$, for each pulsar:
\begin{equation}
  \Omega = -\frac{\dot\nu}{\nu} \frac{n_{obs}(1 + A\cos\varphi) -
n}{A\sin\varphi},
  \label{eq_omega}
\end{equation}

\begin{figure}
  {\centering \resizebox*{1\columnwidth}{!}{\includegraphics[angle=270]
      {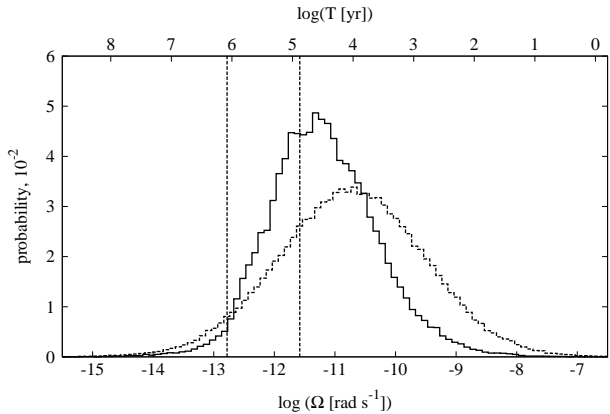}} \par}
  \caption{The distribution of a timescales of long-term variations,
    estimated according to the observed pulsars' frequency and its derivatives,
    for spin-down parameters values derived in Section~\ref{subsec_results}
    (see text for details). The timescales are clustered around $5-500$
    thousands of years and are in a good agreement with similar distribution of
    NS precession timescales caused by anomalous braking torque (dotted line,
    see Section~\ref{subsec_anomal}). Dashed lines represent timescales
    $T_{up}$ and $T_{typ}$ according to estimations (\ref{eq_Tup}) and
(\ref{eq_Ttyp}),
    respectively, for $\dot\nu = 10^{-14}$ $s^{-2}$.}
  \label{fig_omegas}
\end{figure}

To get the distribution of {\it all} possible $\Omega$'s for the observed set
we simulated uniformly distributed phases $\varphi$, as well as values of $n$
and $A$ -- according to probability distributions within the confidence regions
derived in Section~\ref{subsec_results}.  Negative $\Omega$ values were
rejected during this simulation as corresponding to physically impossible
combinations of parameters.  Distribution of $\Omega$ simulated for
$\sigma[A] = 0.1$ is shown in Figure~\ref{fig_omegas}. Corresponding timescales
are clustered inside a $5-500$ thousands of years region.

Also, the Figure~\ref{fig_omegas} displays theoretical distribution for the
timescales of NS precession caused by an anomalous braking torque (see
Section~\ref{subsec_anomal} and equation (\ref{eq_anomal})) for comparison.
These
timescale values depend on a number of physical parameters of NS, such as
surface magnetic field, initial period etc. We used distributions of $B$ and
$P_0$ derived by \citet{fgk06}: $\log B$ and $P_0$ distributed normally with
$\langle \log B [G] \rangle = 12.65 $, $\sigma_{\log B} = 0.55$,
$\langle P_0 \rangle = 0.3$ s and $\sigma_{P_0} = 0.15$ s. Masses and radii of
NS have been taken distributed uniformly within $1.3-1.6$ $M_{\odot}$ and
$8-12$ km intervals, respectively. Initial magnetic dipole angles have
been chosen randomly from $[0;\pi/2]$ interval.

Both obtained distributions are in a good agreement with each other and
consistent with the idea of a long-term thousands-of-years variations of a
pulsars' timing parameters.

\section{Discussion}
\label{sec_discuss}

In this work we revise the problem of anomalous values of isolated radiopulsars
braking indices. Since individual values of $\ddot\nu$ are strongly dominated
by non-monotonous component of the spin-down, we performed a statistical
analysis of the ensemble of 297 objects. As a result, we concluded that
pulsars' behaviour on the $\ddot\nu-\dot\nu$ and $\ddot\nu-\nu$ diagrams
suggest the existence of long-term cyclic mechanism affecting observed timing
parameters. We constructed a semi-phenomenological model of observed pulsar
spin-down and estimated it parameters.

We found that the timescale of the $\ddot\nu$ long-term variations is likely
close to
few thousands of years -- it follows both from simple model-independent
estimation (Section~\ref{sec_limits}) and from accurate maximum likelihood
analysis (Section~\ref{sec_timescales}).


In general, our analysis inevitably suggests that observed $\dot\nu$ of
radiopulsars vary with large enough relative amplitude. As a result, basic
pulsar parameters depending on $\dot\nu$ may be significantly over- or
underestimated. We already said some words about this effect in
Section~\ref{subsec_ntau}, and below we'll continue its discussion.

\subsection{Ages of young pulsars revised}

\begin{table*}
\centering
\caption{Corrections for characteristic ages and magnetic fields of several pulsars due to
  $\dot\nu$ variations. Here $\varepsilon$ is an estimated relative $\dot\nu$
  shift assuming evolutionary braking index $n = 3$ and $\delta\ddot\nu = 0$, while
  $P_0$ is an initial pulsar period that is necessary to provide the same
  correction.}
\begin{tabular}{lcccccccc}
\hline
PSR & $\tau_{ch}$, yr & B, G & $\varepsilon$ & $\tau_{ch,ev}$, yr & $B_{ev}$,
G & SNR & SNR
age, yr &
$P_0$, s \\
\hline
B1853+01 & $2\times10^4$& $7.6\times10^{12}$& -0.50 & $10^4$&
$1.1\times10^{13}$& W44 & $10^4$ [1] & 0.190 \\
B2334+61 & $4\times10^4$& $9.9\times10^{12}$& -0.46 & $2.1\times10^4$ &
$1.3\times10^{13}$& G114.3+0.3
& $(1-2)\times10^4$ [2] & 0.346 \\
B1758-23 & $60\times10^3$ & $7.0\times10^{12}$ & -0.89 & $6.6\times10^3$ &
$2.1\times10^{13}$ & W28 &
$(2.5-150)\times10^3$ [3,4] & 0.391 \\
\hline
{\small [1] \cite{rho94}}
& & & & & & \\
{\small [2] \cite{fich86}}
& & & & & & \\
{\small [3] \cite{lon91}}
& & & & & & \\
{\small [4] \cite{kas93}}
& & & & & & \\
& & & & & & \\
\end{tabular}
\label{tab_psrsages}
\end{table*}

Let's return to the diagram $n_{obs}-\tau_{ch}$ (Fig.~\ref{fig_ntau}). There
are eight pulsars in area III with $n_{obs} > 0$, associated with young
supernova remnants. Assuming their $n_{obs}$ are biased mostly due to $\dot\nu$
variations, i.e. $\eta = 0$, corresponding instant $\varepsilon$ can be
derived as 
\begin{equation}
\varepsilon =  \sqrt{3/n_{obs}} - 1
\end{equation} 
Therefore, their evolutionary characteristic ages
\begin{equation}
  \label{eq_tau_ev}
  \tau_{ch,ev} = \tau_{ch}(1 + \varepsilon)
\end{equation}
and magnetic fields
\begin{equation}
  B_{ev} = \frac{B}{\sqrt{1 + \varepsilon}},
\end{equation}
where $B$ is according to equation (\ref{eq_B}), can be estimated, which are
expected to be closer to the physical ones.

We analysed data on three of these eight supernova remnants that have more or
less independent and confident estimations of their ages, and summarized the
results in Table~\ref{tab_psrsages}.  
In general, corrected pulsars' ages become more consistent with
that of corresponding SNRs in all of these cases. While the estimations of
magnetic fields in all cases exceed $10^{13}$ G. At the same time, if
one assume initial periods $P_0$ for these pulsars, then the same
corrections of characteristic ages can be provided by the proximity
of the $P_0$ and the current period.

A similar correction for the pulsars of area I is not so evident.  Whether the
measured $n_{obs} < 3$ of the youngest Crab-like pulsars are indeed biased from
evolutionary $n = 3$ due to the same variational process is not clear. However,
according to equation (\ref{eq_tauobs}), pulsars born with nonzero
characteristic ages
$P_0/2K \sim 10^3$ years. If so, then all pulsars in area I likely have
$\tau_{ch,ev}$ greater than at least 2 thousands of years while their measured
$\tau_{ch} < 2$ kyr imply positive $\varepsilon$.

We believe that understanding of a possibility of significant biasing of
standard estimators of pulsars' ages will induce new insights on the PSR-SNR
connections, evolution and kinematics.

\subsection{Long timescale precession of a neutron star?}
\label{subsec_anomal}

And finally we will discuss possible physical nature of a long-term variations
of observed pulsars' spin-down suggested in this work.

The unusual timescale of these variations -- thousands of years -- is much
larger than the pulsar rotational period and, on the other hand, much shorter
than its typical lifetime. At the same time, even the canonical magnetodipolar
braking model seems to already include an essential mechanism for long-term
variations.

More than half a century ago \citet{deu55} derived the vacuum solution for the
electromagnetic field of a perfectly conducting, rigidly rotating spherical
star. The full braking torque affecting such a star with magnetic moment
$\vec \mu$ and spin angular velocity $\vec\omega$ may be described as a
superposition of three orthogonal terms (in observer's frame):
\begin{equation}
  \label{eq_torque}
  \dot{\vec\omega} = \alpha \cdot (\vec\mu \times \vec \omega) + \beta
  \cdot ((\vec\mu  \times \vec\omega) \times \vec\omega) + \gamma \cdot
  \vec\omega
\end{equation}
where $\mu = const$. The second and third components of this braking
torque are due to the magnetodipolar radiation in far (wave) zone. Both of
them vary as $(\omega R/c)^3$, where $R$ is the star radius \citep{dav70}. At
the same time, first component of torque (\ref{eq_torque}) is due to radiation
in near
zone and varies as $(\omega R/c)^2$. Since $\omega R/c \approx
10^{-4}-10^{-2}$ for a typical pulsar, the near-field torque is up to
four orders of magnitude greater than far-field one.

However, being very strong, this torque does not directly affect pulsar
spin-down rate and can not change magnetic inclination angle $\chi$, as it is
always perpendicular to both rotational and magnetic axes, in contrast to the
far-field part of the torque driving the evolution of pulsar' $\omega$ and
$\chi$ \citep{dav70}.

This near-field torque induces an additional complex
rotation of the star. Indeed, first term of torque (\ref{eq_torque}) describes a
precession of a neutron star around its magnetic axis with angular frequency
$\Omega = |\alpha\mu|$. Due to the strength of the near-field torque, the
characteristic timescale of this precession is significantly shorter than the
pulsars' typical lifetime. Precisely, in a simple vacuum magnetodipolar case,
precession period remains constant and is equal to \citep{gn85}
\begin{equation}
  T = \frac{2\pi}{\Omega} = (2.9 \times 10^3\mbox{ yr}) \cdot I_{45} \cdot
  \nu^{-1}_{i,50} \cdot B^{-2}_{s,12} \cdot R^{-5}_{6} \cdot
  \cos^{-1}\chi_0
  \label{eq_anomal}
\end{equation}
Here $\nu_i$ -- pulsar's initial frequency normalized to $50\mbox{ Hz}$;
$B_{s,12}$ -- surface magnetic field normalized to $10^{12}\mbox{ G}$; $I_{45}$
-- NS moment of inertia normalized to $10^{45}\mbox{ g}\cdot\mbox{cm}^2$, and
$R$ -- NS radius normalized to $10^6\mbox{ cm}$; $\chi_0$ -- initial magnetic
inclination angle. The value of $\Omega \ll \omega$ depends strongly on the NS
parameters, hence one can expect a wide range of precession periods, from
hundreds to thousands years. The distribution of these periods for reasonable
values of NS parameters is shown in Figure~\ref{fig_omegas} and is in a good
agreement with estimations of variational process timescales derived in this
work.

Note, that similar long-term precession on a timescales of few thousand years
can be also caused by the NS distortion along the magnetic axis due to the
strong magnetic field \citep{gol70}.

However, if we associate pulsar beam symmetry axis with NS magnetic moment then
it is possible to show that such NS precession itself can not cause large observed
deviations of $\ddot\nu$ (and $\dot\nu$) -- moreover, it is unable to power a
cyclic variations in observed NS spin-down rate at all.  Indeed, the braking
torque vector (\ref{eq_torque}) is dominated by the strong near-field
term. Hence, assuming $\beta \approx 0$ and $\gamma \approx 0$:
\begin{equation}
  \label{eq_dotomega}
  \dot{\vec\omega} \approx \alpha\cdot(\vec\mu \times \vec\omega) = (\vec\Omega
\times \vec\omega),
\end{equation}
where $\Omega $ is a constant precession frequency. At the same time, observed
pulsar spin frequency is defined by rotation of $\vec \mu$. However, this
vector rotates around angular velocity $\vec\omega$:
\begin{equation}
  \label{eq_dotmu}
  \dot{\vec\mu} = (\vec\omega \times \vec\mu),
\end{equation}
which also changes its direction relative to observer. Therefore, the value of
{\it observed} spin frequency of $\vec\mu$ is not equal to $\omega$ and is not
clear from equations (\ref{eq_dotomega}) and (\ref{eq_dotmu}) only.

But, combining these two equations we found that
$\dot{\vec\omega} + \alpha\dot{\vec\mu} = 0$, which suggests introduction of
the vector
\begin{equation}
\vec L = \vec\omega + \alpha\vec\mu = \vec\omega + \vec\Omega = const
\end{equation}
which is useful for the description of $\vec\mu$ rotation by a simpler
equation similar to (\ref{eq_dotmu}):
\begin{equation}
  \dot{\vec\mu} = (\vec L \times \vec\mu)
\end{equation}
Since $\vec L = const$, it is clear that a star precessing around its magnetic
axis looks like a non-precessing star with a bit biased observed spin frequency
\begin{equation}
  L \approx \omega + \Omega\cos\chi,
\end{equation}
where $\chi$ is a magnetic inclination angle.

Therefore, even slow {\it monotonous} evolution of $\Omega$, $\omega$ and $\chi$
are unable to introduce any irregularities in observed spin-down rate. The
discussed type of precession is indiscernible for an observer.

Therefore, if observed long-term spin-down variations are really caused by such
precession, some geometrical or physical mechanism connecting rotation of
$\vec\omega$ around $\vec\mu$ with modulation of $\chi$, $\dot\omega$ etc
should be suggested.

Thus, \cite{bar10} considered an alteration of electric currents in the NS
magnetosphere caused by such precession, which leads to modulation of
$\dot\omega$. At the same time, \citet{mel00} have shown that for non-spherical
but biaxial or triaxial NS, near-field torque will strongly affect its
rotation. If pulsar's rotational axis (and magnetic moment) significantly
deviates from the one of its main inertia axes, then a very large variations of
$\chi$ (and consequently of $\dot\omega$) occur.  However, variations so large
are not observed \citep{mel00}.

Generally, any change in the observed pulsar timing parameters are determined
by the variations of the $d\psi/dt$, where $\psi$ is the dihedral angle between
two planes -- the plane of symmetry axis of a pulsar beam and the plane formed
by rotational axis and direction to the observer. Detailed investigations of
observed complex timing behaviour should always take into account the geometry of
pulsar emission.

Ultimately, whether the monotonous precession of NS rotation axis around the
magnetic moment able to produce any non-monotonous peculiarities in observed
$\dot\nu$ and $\ddot\nu$ is not clear, but at least this process operates on
sufficiently long timescales of thousands of years.  On the other hand, there
are no obvious physical reason behind some hypothetical stochastic process with
ultra-red spectrum, able to produce the same amplitude of pulsars' second
derivatives variations. The regular nature of these variations seems
phenomenologically preferable.

\section{Conclusions}
\label{sec_conclude}

In this work we analysed the properties of a set of 297 ``ordinary'' radio
pulsars
with a well detected, via a pulsar timing, second frequency
derivative. As a result we demonstrate that

\begin{itemize}
\item the long term spin-down of pulsars is well described by the superposition
of
  a ``true'' monotonous and a long timescale non-monotonous cyclic term;
\item the subsets of these pulsars with positive and negative second
  derivatives are statistically different, both in shape and in number of
  objects; this effect is stronger for younger pulsars and vanishes
  for the older ones;
\item this difference reflects the existence of an evolutionary, secular
  positive trend $\ddot\nu_{ev}$; the $\ddot\nu$ variations around this
  trend are nearly symmetric;
\item simple binomial arguments based on the assumption of symmetric variations
  around the evolutionary trend allow to construct reliable estimator for
  pulsars' long-term spin-down parameters;
\item pulsars secular spin down is consistent with a classical
  magnetodipolar braking with $n\approx3$
\item the mean relative amplitude of the first frequency derivative ($\dot\nu$)
  variations is as large as $\langle A\rangle = 0.5-0.8$ with variance
$\sigma[A]\approx0.1$
\item the characteristic timescale of these variations is likely to be of
  several thousands years;
\item consequently, the observed values of pulsars' characteristic age are
  biased by the factor of $0.5-5$ from the ``true'' secular value
  $\tau_{ch,ev}$; this fact naturally explains the discrepancy between real
  ages and characteristic ages of several pulsars associated with supernova
  remnants, the presence of pulsars with extremely large, up to $\sim 10^8$
  years, characteristic ages, and low braking indices of the youngest pulsars.
\item there is at least one physical reason able to produce such variations
  in magnetodipolar model -- a complex neutron star rotation relative its
  magnetic axis due to influence of
  the near-field part of magnetodipolar torque. 
\end{itemize}

\section*{Acknowledgments}
This work has been supported by the Russian Foundation for Basic Research
(Grant No. 04-02-17555), Russian Academy of Sciences (program ``Evolution of
Stars and Galaxies''), by the Russian Science Support Foundation, by the Grant
of President of Russian Federation for the support of young Russian scientists
(MK-4694.2009.2) and by the grant of Dynasty foundation. The authors are
grateful to the anonymous Referee for his/her thoughtful review which lead to
significant improvement of the paper.


\label{lastpage}

\end{document}